\documentclass[10pt,notitlepage,a4paper,aps,prd,onecolumn,superscriptaddress,nofootinbib,groupedaddress]{revtex4-1}

\usepackage{bm}
\usepackage{array}

\usepackage[dvipsnames]{xcolor}
\usepackage{graphicx}
\usepackage{subcaption}
\usepackage{amsthm}

\theoremstyle{plain}

\theoremstyle{remark}

\usepackage[pdfencoding=auto, psdextra]{hyperref}
\usepackage[T1]{fontenc}
\usepackage[utf8]{inputenc}
\usepackage[english]{babel}
\usepackage{hyperref}
\usepackage[mathscr]{euscript}
\usepackage{caption}
\usepackage{subcaption}
\hypersetup{
colorlinks = true,
linkcolor = blue,
filecolor = magenta,
urlcolor = blue,
pdftitle = {TLSs interaction},
pdfpagemode = FullScreen,
citecolor = blue}
\usepackage{physics}

\usepackage{caption}

\usepackage{indentfirst}

\usepackage{amsmath}
\allowdisplaybreaks  % para que pueda partir fórmulas que ocupan más de una línea, necesita el paquete anterior
\usepackage{amssymb} % para cargar algunos símbolos como \blacksquare y \square
\usepackage{amsfonts} % para cargar algunas fuentes en estilo matemático
\usepackage{enumerate}

\usepackage{fancyhdr}                     % para definir distintos tipos de cabeceras y pies de página

\usepackage{float}                             

\usepackage{amsmath}         % For advanced math typesetting
\usepackage{amsfonts}        % For additional mathematical fonts
\usepackage{amssymb}         % For more math symbols
\usepackage{amsthm}          % For theorem environments
\usepackage{empheq}          % For enhanced equation environments

\usepackage{color}           % For colored text
\usepackage{ragged2e}        % For ragged right alignment
\usepackage{titlesec}        % For customizing section title formats

\usepackage{tipa}            % For the thorn symbol
\usepackage[utf8]{inputenc} % For UTF-8 input encoding
\allowdisplaybreaks          % Allow page breaks within equations

\usepackage{enumitem}        % For customizable lists

\usepackage{tikz}            % For creating graphics
\usetikzlibrary{shapes.geometric,arrows.meta,arrows,positioning}  % Load specific TikZ libraries
\usetikzlibrary{tikzmark}   % For advanced TikZ marking

\titleformat{\paragraph}
  {\normalfont\normalsize\bfseries}{\theparagraph}{1em}{} % Format paragraph titles
\titlespacing*{\paragraph}
  {0pt}{3.25ex plus 1ex minus .2ex}{1.5ex plus .2ex} % Set spacing for paragraph titles

\begin{document}
\title{Unstable Periodic-Orbit Localization in a Nonlinear System--Environment Model}

\author{Daniel Martínez-Gil}
\email{daniel.martinez@upv.es}
\affiliation{Universitat Politècnica de València, Camino de Vera s/n, 46022, Valencia, Spain}

\begin{abstract}The role of unstable periodic orbits (UPOs) is investigated in the chaotic dynamics of a nonlinear system--environment model composed of a central two-level system coupled to an environment of two-level systems. Families of UPOs embedded in chaotic regions are identified and their local instability is characterized. Their behavior is examined under variations of the environment size, energy, and system symmetry, allowing the model to serve as a controlled testbed in which the dimensionality and dynamical conditions can be systematically modified.

Characteristic UPO-based dynamical quantities are used to assess the influence of these periodic orbits on nearby chaotic trajectories, including occupation of the UPO tube, its persistence at late times, return fidelity enhancement, and recurrence timing. Across the different model variations considered, trajectories initialized close to the UPOs retain statistically detectable spatial and temporal signatures associated with the periodic orbits. These results indicate that unstable periodic orbits can retain a measurable organizing influence on surrounding chaotic trajectories across changes in dimensionality, energy, and system symmetry, giving rise to scar-like dynamical signatures in a simple nonlinear system--environment model.
\end{abstract}

\maketitle

\section{Introduction}

Chaotic dynamical systems are characterized by a strong sensitivity to
initial conditions: trajectories starting arbitrarily close to one another
may eventually evolve in very different ways. Despite this apparent
irregularity, chaotic dynamics contains recurrent structures that can provide
a geometrical description of the underlying phase-space organization
\cite{EckmannRuelle1985,Auerbach1987,Cvitanovic1991,Cvitanovic1992}.
Among these structures, unstable periodic orbits (UPOs) are trajectories
that return exactly to their initial state after a finite period $T$, while
remaining unstable against perturbations. An initial condition placed
exactly on a UPO repeats the same motion indefinitely, whereas a nearby
trajectory generally follows a similar evolution only temporarily before
separating from it. This local influence suggests that UPOs may affect the
surrounding chaotic dynamics even after nearby trajectories begin to separate
from the periodic path. Collections of UPOs can therefore provide a useful
skeleton for describing chaotic dynamics and, in suitable systems, their
statistical properties
\cite{Auerbach1987,Cvitanovic1991,Cvitanovic1992}.

UPOs are particularly relevant in quantum physics through their connection with quantum scars. In its original single-particle formulation, quantum scarring refers to quantum states that become unusually concentrated around unstable periodic orbits of their classical counterparts. Heller first identified this behavior in eigenfunctions of chaotic Hamiltonian systems, and subsequent work established connections between scar strength, recurrence, and the instability of the underlying periodic orbit
\cite{Heller1984,Bogomolny1988,Berry1989,KaplanHeller1999}. The concept has since been extended to many-body systems, where atypical nonthermal states and persistent dynamical revivals can appear within otherwise ergodic spectra
\cite{Turner2018,Serbyn2021,Moudgalya2022,Chandran2023}. Closely related phenomena have also been studied in the Dicke model, coupled-spin systems, chaotic mean-field dynamics, and interacting bosonic systems, where unstable or recurrent classical or mean-field motion has been connected with localization, dynamical revivals and persistent memory
\cite{PilatowskyCameo2021,PilatowskyCameo2022,Mondal2020,Hummel2023,
EvrardSpinor2024,Evrard2024,Pizzi2025,Ermakov2025}.

The periodic-orbit viewpoint also has applications well beyond quantum physics. A particularly developed example is fluid turbulence, where the flow can be represented as a trajectory through a very high-dimensional state space that repeatedly approaches unstable invariant solutions. Periodic and relative periodic solutions have been identified in shear flows and two-dimensional turbulence, and increasingly large catalogues of recurrent solutions have been used to characterize both the geometry and the statistics of turbulent motion
\cite{KawaharaKida2001,Viswanath2007,KawaharaUhlmannVanVeen2012,
ChandlerKerswell2013,LucasKerswell2015,GrahamFloryan2021,Yalniz2021,
Page2024,Ashtari2026}. Related UPO-based descriptions have also been applied to other nonlinear problems, including neuronal dynamics, nonlinear fluid systems, atmospheric regime transitions, and the stabilization and control of chaotic systems
\cite{So1998,EncisoLuquePeralta2020,EncisoPeraltaRomaniega2023,
OttGrebogiYorke1990,Nieto2022,Hamilton2025}.These examples illustrate the broader role of unstable recurrent structures in the analysis of complex nonlinear dynamics.

Periodic-orbit methods form part of a broader set of tools for characterizing deterministic chaos. Alternative approaches based on symbolic dynamics, ordinal patterns, permutation entropy, complexity measures, and mathematical analyses of nonlinear evolution equations provide complementary ways of identifying structure in complex dynamical systems
\cite{BandtPompe2002,AmigoKennelKocarev2005,Amigo2006,Amigo2010,
Makarov2023,Conejero2016}.

The model studied here shares several dynamical ingredients with these
UPO-related scar studies. It possesses unstable periodic trajectories
embedded in chaotic regions, and it is examined whether trajectories initialized
close to them retain enhanced spatial localization and recurrent dynamics
associated with their characteristic periods. There is, however, an
important conceptual distinction. The present system is governed by
nonlinear, state-dependent system--environment equations rather than by the
eigenstates of a fixed linear quantum Hamiltonian. Consequently, the
behavior investigated here does not satisfy the conventional definition of
Heller-type quantum scarring. Therefore, the phenomenon studied below is referred to as \emph{UPO-associated dynamical localization}. Nevertheless, the
similarities in periodic-orbit localization, recurrence, and the competition
between orbital instability and recurrence timescale provide a direct point
of contact with the scar-related literature
\cite{KaplanHeller1999,PilatowskyCameo2021,Mondal2020,Evrard2024,Pizzi2025}.

The model originates from the geometry-based formulation of interacting TLS
dynamics introduced in Ref.~\cite{MartinezGil2025} and subsequently applied to a chiral system--environment setting in Ref.~\cite{MartinezGil2025Chiral}. Each TLS is
represented through variables associated with its Bloch-sphere geometry,
while the interaction depends on the population imbalance of the coupled
subsystems. The resulting nonlinear equations have a mathematical form
reminiscent of Gross--Pitaevskii-type dynamics, although their physical
origin and interpretation are different. Previous work focused primarily
on the resulting system--environment evolution and the emergence of
oscillatory, tunneling-suppressed, and dissipative-like regimes
\cite{MartinezGil2025}. Here, the same comparatively simple model is used
as a controlled setting for studying how unstable periodic motion can
organize nonlinear chaotic dynamics.

Homogeneous environments containing
$N=2,4,6,8,$ and $10$ TLSs at a common energy density are first investigated. The chaotic background is characterized using finite-time Lyapunov exponents (FTLEs)
\cite{Benettin1980}, while the local instability of the identified periodic
orbits is quantified through a fixed-sector Floquet analysis that respects
the invariant structure of the homogeneous dynamics. As described in
Sec.~\ref{sec:dynamical_localization}, complementary spatial and temporal metrics are
then used to quantify the influence of these UPOs on the surrounding chaotic
dynamics.
The representative case \(N=10\) is also considered at five different energy densities, for which the complete analysis is repeated in order to examine how the observed behavior changes across different dynamical regimes.

Finally, the analysis is repeated for two additional cases. In the first, a finite energetic bias is introduced in the central TLS while the environment remains homogeneous. In the second, different energetic biases
are assigned to the environmental TLSs, breaking the environmental
homogeneity. These calculations are again performed for $N=2,4,6,8,$ and
$10$ at the reference energy density and test whether the observed
UPO-associated dynamical signatures survive as the symmetry and collective
structure of the reference model are progressively removed.

The purpose of the present work is therefore broader than the particular
TLS realization. The model provides a compact nonlinear testbed in which
the number of environmental degrees of freedom, energy, symmetry, and
collective structure can be varied independently while retaining direct
access to unstable periodic orbits and the surrounding chaotic dynamics.
Because the formulation is expressed in terms of generic Bloch-vector
degrees of freedom and state-dependent couplings, the results apply
directly to systems governed by the same nonlinear equations rather than to
a single specific physical implementation. More generally, the methodology developed here can be transferred to other nonlinear dynamical systems in which unstable recurrent structures coexist with chaotic motion. The established role of such structures in
areas ranging from turbulence and geophysical dynamics to neuronal systems
and chaos control further illustrates the broader relevance of this
dynamical-systems viewpoint
\cite{KawaharaUhlmannVanVeen2012,GrahamFloryan2021,Page2024,
Hamilton2025,So1998,OttGrebogiYorke1990}. 

This manuscript is structured as follows. Section~\ref{sec:model} introduces the nonlinear system--environment model and its dynamical formulation. Section~\ref{sec:upos} characterizes the chaotic dynamics, identifies the unstable periodic orbits considered in the subsequent analysis, and quantifies their local instability through Floquet analysis. Section ~\ref{sec:dynamical_localization} examines the influence of these UPOs on nearby chaotic trajectories through spatial and temporal dynamical signatures. The dependence on energy is analyzed in Section ~\ref{sec:energy_dependence}, while Section ~\ref{sec:robustness} investigates the robustness of the observed behavior under symmetry-breaking modifications of the model. Finally, the main conclusions are summarized in Section ~\ref{sec:conclusions}.
% ============================================================
\section{Nonlinear system--environment model}
\label{sec:model}
% ============================================================

The nonlinear system--environment model considered here is based on the
geometry-based representation of two-level systems introduced in
Ref.~\cite{MartinezGil2025}. In this formulation, the
quantum state of each TLS is mapped onto the Bloch sphere after removing its
physically irrelevant global phase. The resulting Bloch coordinates can be
expressed as expectation values of the Pauli operators, providing a
classical-like geometrical representation of the TLS dynamics. In
particular, the population difference and relative phase form a pair of
canonically conjugate variables. A detailed derivation of this mapping is given in
Ref.~\cite{MartinezGil2025}.

A central TLS coupled to an environment formed by \(N\) additional TLSs is
considered. Each environmental TLS interacts with the central TLS, while no
direct interaction is introduced between different environmental TLSs.

The central TLS is represented by the Bloch vector
\begin{equation}
    \bm{r}_0=(x_0,y_0,z_0),
\end{equation}
and the $i$th environmental TLS by
\begin{equation}
    \bm{r}_i=(x_i,y_i,z_i),
    \qquad i=1,\ldots,N.
\end{equation}
For pure states,
\begin{equation}
    |\bm{r}_0|=|\bm{r}_i|=1.
\end{equation}
Here, $z_0$ and $z_i$ denote the corresponding population differences,
while $\Phi$ and $\phi_i$ denote the relative phases of the populations. The Bloch-sphere
coordinates can therefore be written as
\begin{align}
    x_0 &= \sqrt{1-z_0^2}\cos\Phi,
    &
    y_0 &= \sqrt{1-z_0^2}\sin\Phi,
    \\
    x_i &= \sqrt{1-z_i^2}\cos\phi_i,
    &
    y_i &= \sqrt{1-z_i^2}\sin\phi_i.
\end{align}
Throughout the manuscript, the subscript $0$ denotes the central TLS, whereas
the index $i=1,\ldots,N$ labels the TLSs forming the environment.

In units with $\hbar=1$, the Hamiltonian function is written as
\begin{equation}
    H=H_{\mathrm S}+H_{\mathrm E}+H_{\mathrm I},
\end{equation}
with
\begin{equation}
    H_{\mathrm S}
    =
    -2\delta\sqrt{1-z_0^2}\cos\Phi
    +2\epsilon z_0 ,
\end{equation}
\begin{equation}
    H_{\mathrm E}
    =
    \sum_{i=1}^{N}
    \left[
        -2\delta_i\sqrt{1-z_i^2}\cos\phi_i
        +2\epsilon_i z_i
    \right],
\end{equation}
and
\begin{equation}
    H_{\mathrm I}
    =
    z_0\sum_{i=1}^{N}\Lambda_i z_i,
    \label{eq:interaction}
\end{equation}
representing the system, the environment, and the interaction Hamiltonian, respectively.
Here, $\delta$ and $\delta_i$ control the tunneling dynamics of the central
and environmental TLSs, respectively, $\epsilon$ and $\epsilon_i$ describe
their energetic asymmetries, and $\Lambda_i$ is the coupling strength between
the central TLS and the $i$th environmental TLS.

In this representation, the population difference
plays the role of a generalized momentum. The interaction in
Eq.~\eqref{eq:interaction} can therefore be interpreted as a
momentum--momentum-type coupling. A detailed discussion of this construction
and of alternative bilinear couplings is given in
Ref.~\cite{MartinezGil2025}.

An important feature of the present model is that the interaction differs
from the conventional spin--spin coupling of a linear many-spin
Hamiltonian. In the quantum representation,
\begin{equation}
    H_{\mathrm I}
    =
    \langle\hat{\sigma}_z\rangle_0
    \sum_{i=1}^{N}
    \Lambda_i
    \langle\hat{\sigma}_z\rangle_i ,
\end{equation}
so that the interaction is constructed from products of expectation values
rather than from an operator product such as
$\hat{\sigma}_z^{(0)}\hat{\sigma}_z^{(i)}$. The resulting evolution is
therefore nonlinear and explicitly dependent on the instantaneous state of
the interacting subsystems \cite{MartinezGil2025}.

A homogeneous environment is first considered,
\begin{equation}
    \delta_i=\delta_{\mathrm e},
    \qquad
    \epsilon_i=\epsilon_{\mathrm e},
    \qquad
    \Lambda_i=\Lambda ,
\end{equation}
and use the parameters
\begin{equation}
    \delta=\delta_{\mathrm e}=1,
    \qquad
    \epsilon=\epsilon_{\mathrm e}=0,
    \qquad
    \Lambda=1,
\end{equation}
with $\delta$ setting the energy scale. In this homogeneous case, the
identical environmental TLSs also allow an exact collective description of
the dynamics. The derivation and the associated conserved quantities are
given in Appendix~\ref{app:collective_structure}; only their consequences
when constructing dynamically comparable trajectories will be needed below.

To compare systems containing different numbers of environmental TLSs, the energy per TLS is kept fixed,
\begin{equation}
    e=
    \frac{\mathcal{E}_N}{N+1}
    =
    0.83711739,
    \label{eq:fixed_energy_density}
\end{equation}
where $\mathcal{E}_N$ is the conserved total energy, i.e. the value of the
Hamiltonian $H$ for a system formed by the central TLS and $N$
environmental TLSs. The denominator $N+1$ is therefore the total number of
TLSs in the system. The following environment sizes are considered:
\begin{equation}
    N=2,4,6,8,10.
\end{equation}
Keeping \(e\) fixed, rather than the total energy \(\mathcal{E}_N\), allows systems with different numbers of TLSs to be compared at the same energy per TLS.

The homogeneous system defined above is used as the reference model. In the following section, it is first determined whether the corresponding energy regions display chaotic dynamics, after which unstable periodic orbits embedded in them are searched for.

% ============================================================
\section{Chaotic dynamics and unstable periodic orbits}
\label{sec:upos}
% ============================================================

The chaotic dynamics of the homogeneous model are next characterized, and the periodic orbits subsequently used to test UPO-associated dynamical localization are identified.

% ------------------------------------------------------------
\subsection{Chaotic dynamics at fixed energy density}
\label{subsec:chaos}
% ------------------------------------------------------------

As defined in Sec.~\ref{sec:model}, environments with
\begin{equation}
    N=2,4,6,8,10
\end{equation}
are compared at the same energy per TLS,
\begin{equation}
    e=0.83711739.
\end{equation}
The purpose of varying $N$ is to determine how the chaotic dynamics changes
as the finite environment becomes larger. It is necessary to remark at this point that these values are not intended to
represent a thermodynamic limit. Instead, they provide a controlled finite-size sequence in which the number
of environmental degrees of freedom is progressively increased while
keeping the remaining model parameters fixed.

A direct evaluation of the accessible energy ranges confirms that the chosen
energy corresponds to comparable regions of the available energy range for
the five environment sizes. The accessible energy intervals and the relative
position of the selected energy within them are discussed in
Appendix~\ref{app:energy_ranges}.

Before searching for periodic orbits, it is first verified that the selected energy regions exhibit sensitivity to initial conditions. This is quantified using finite-time Lyapunov exponents (FTLEs) \cite{Benettin1980}.

The basic idea is to compare two trajectories whose initial conditions differ
by a very small amount. If their separation is denoted by $d(t)$, in a
chaotic region it grows approximately as
\begin{equation}
    d(t)
    \sim
    d(0)e^{\lambda t},
\end{equation}
where $\lambda$ is the Lyapunov exponent. A positive value therefore means
that initially close trajectories separate exponentially with time and
provides a direct measure of sensitivity to initial conditions.

In the present calculations, the exponent is evaluated over a finite
evolution time, giving the finite-time Lyapunov exponent
$\lambda_{\mathrm{FT}}$. The calculation starts from a reference trajectory
and a nearby perturbed trajectory separated by a small distance $d_0$.
Both trajectories are then evolved for a short time interval $\Delta t$.
At the end of the first interval, their separation has changed from $d_0$
to $d_1$, and the corresponding growth factor is recorded.

Before starting the next interval, the perturbed trajectory is redefined so
that its distance from the reference trajectory is again $d_0$. Its
direction relative to the reference trajectory is kept equal to the
direction of the separation reached at the end of the previous interval.
The two trajectories are then evolved for another interval $\Delta t$, at
the end of which a new separation $d_2$ is obtained. Repeating this procedure
produces a sequence of separations $d_k$, where $k$ labels the successive
evolution intervals.

The finite-time Lyapunov exponent is then calculated from the accumulated
growth of these separations,
\begin{equation}
    \lambda_{\mathrm{FT}}
    =
    \frac{1}{t_{\mathrm{tot}}}
    \sum_k
    \ln\left(\frac{d_k}{d_0}\right).
\end{equation}
A positive value indicates that nearby trajectories tend to separate
exponentially over the considered time interval.
All trajectories considered in the same-energy ensembles yield positive
values of $\lambda_{\mathrm{FT}}$. The median values are
\begin{equation}
    \widetilde{\lambda}_{\mathrm{FT}}
    =
    \{
    0.0266,\,
    0.0256,\,
    0.0706,\,
    0.1830,\,
    0.2355
    \}
    \label{eq:median_ftle_values}
\end{equation}
for $N=2,4,6,8,10$, respectively. The median is used rather than the mean because the FTLE distributions are not necessarily symmetric and may contain tails toward larger values. The median therefore provides a representative
measure of the typical chaotic sensitivity that is less affected by a small
number of unusually large FTLE values.

\begin{figure*}[h]
    \centering

    \begin{subfigure}[t]{0.31\textwidth}
        \centering
        \includegraphics[width=\textwidth]{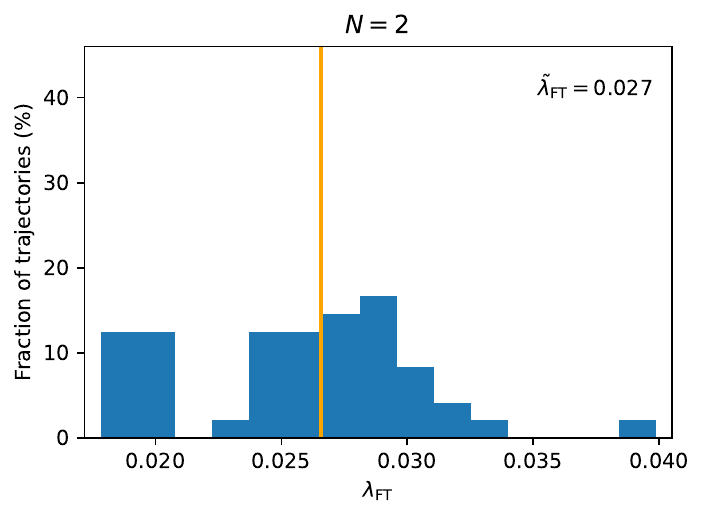}
    \end{subfigure}
    \hfill
    \begin{subfigure}[t]{0.31\textwidth}
        \centering
        \includegraphics[width=\textwidth]{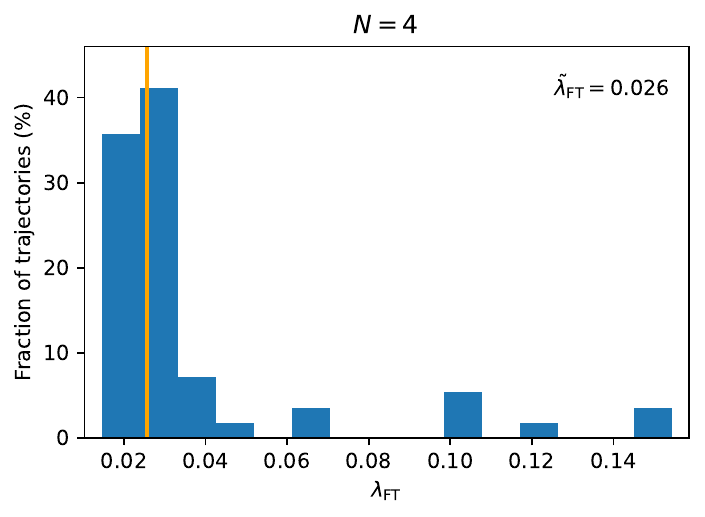}
    \end{subfigure}
    \hfill
    \begin{subfigure}[t]{0.31\textwidth}
        \centering
        \includegraphics[width=\textwidth]{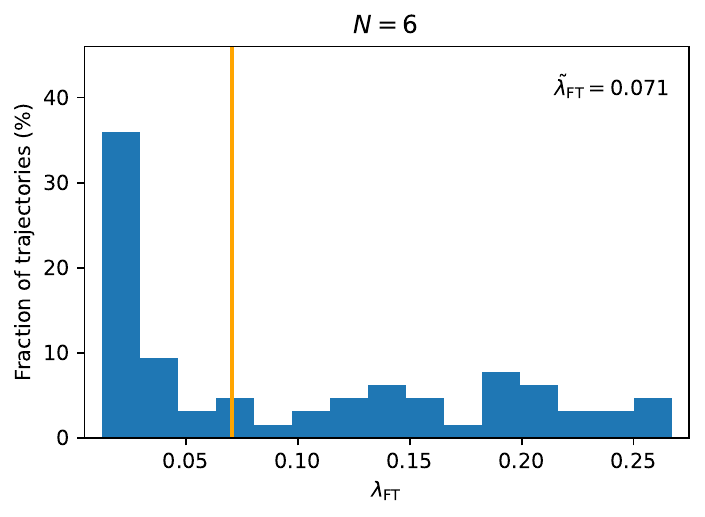}
    \end{subfigure}

    \vspace{0.5em}

    \begin{subfigure}[t]{0.31\textwidth}
        \centering
        \includegraphics[width=\textwidth]{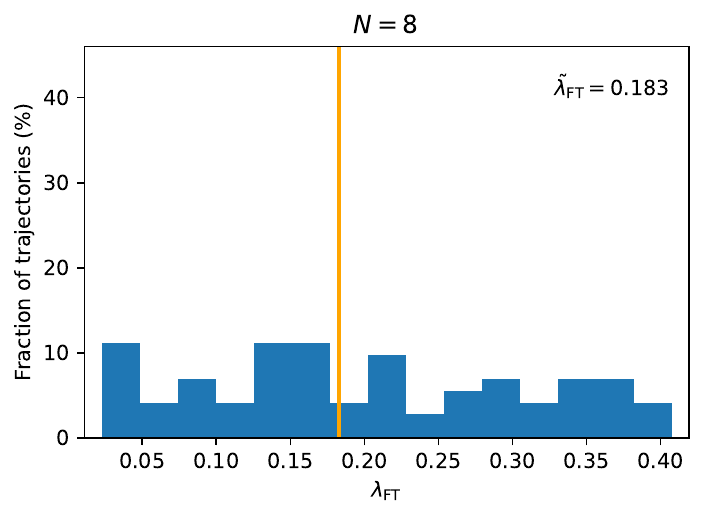}
    \end{subfigure}
    \hspace{0.04\textwidth}
    \begin{subfigure}[t]{0.31\textwidth}
        \centering
        \includegraphics[width=\textwidth]{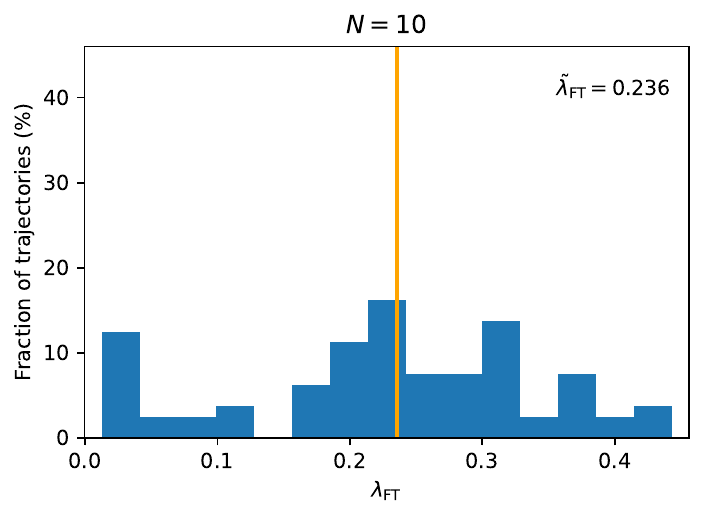}
    \end{subfigure}

    \caption{\justifying
    Distributions of the finite-time Lyapunov exponent
    $\lambda_{\mathrm{FT}}$ for $N=2,4,6,8,$ and $10$ at the common
    energy density used in the analysis. The horizontal axis gives the
    FTLE obtained for each sampled trajectory, while the vertical axis gives
    the fraction of sampled trajectories whose FTLE lies within each histogram
    interval. The orange vertical line marks the median value,
    $\widetilde{\lambda}_{\mathrm{FT}}$, for each distribution.
    }
    \label{fig:ftle_distributions}
\end{figure*}

As shown in Fig.~\ref{fig:ftle_distributions}, the distributions for
$N=2$ and $N=4$ are concentrated at relatively small FTLE values. From
$N=6$ onward, the distributions progressively shift toward larger values,
with the strongest chaotic sensitivity observed for $N=8$ and $N=10$.
The selected energy regions therefore provide a chaotic background in which
to search for unstable periodic orbits.

% ------------------------------------------------------------
\subsection{Search for unstable periodic orbits}
\label{subsec:upo_search}
% ------------------------------------------------------------

Having established chaotic behavior at the selected energies, periodic trajectories embedded in these chaotic regions are subsequently searched for. A periodic
trajectory must satisfy a much more restrictive condition than an arbitrary
trajectory: after a finite time $T$, the complete dynamical state must return
to its initial value.

The complete state is denoted by
\begin{equation}
    \bm{R}
    =
    \left(
    \bm{r}_0,\bm{r}_1,\ldots,\bm{r}_N
    \right).
\end{equation}
A periodic orbit therefore satisfies
\begin{equation}
    \bm{R}(T)=\bm{R}(0).
    \label{eq:periodicity_condition}
\end{equation}
This condition requires not only the central TLS but also every individual
environmental TLS to return to its initial state after one period. It is
therefore the complete system--environment configuration that must close
periodically, rather than only a reduced set of collective coordinates or a
low-dimensional projection of the trajectory.

A randomly chosen chaotic trajectory will generally not satisfy this exact
condition. The numerical procedure used to locate periodic-orbit candidates,
refine them into closed trajectories, and validate their periodicity is
described in Appendix~\ref{app:upo_search}.

For each environment size four distinct periodic orbits are retained, giving a
total of
\begin{equation}
    N_{\mathrm{UPO}}=20
\end{equation}
selected trajectories. Using the same number of orbits for every $N$ ensures
that each environment size contributes equally to the subsequent analysis.

Considering the 20 selected periodic orbits, the fundamental period spans
the range
\begin{equation}
    6.13 \lesssim T \lesssim 112.64.
\end{equation}
The analysis therefore includes periodic trajectories with substantially
different recurrence times.

For visualization, each periodic orbit is represented through the Bloch trajectory of the central TLS, using the coordinates
\begin{equation}
    (x_0,y_0,z_0).
\end{equation}
This representation provides a direct and physically transparent
visualization of the recurrent motion of the central subsystem. 

It should be emphasized that the periodicity criterion is not established
from this visualization alone. The actual UPO condition is always verified
for the complete dynamical state through
Eq.~\eqref{eq:periodicity_condition}. The central Bloch representation is
therefore used only as a convenient way to display representative periodic
trajectories.

\begin{figure*}[h]
    \centering

    \begin{minipage}[t]{0.32\textwidth}
        \centering
        \includegraphics[width=\textwidth]{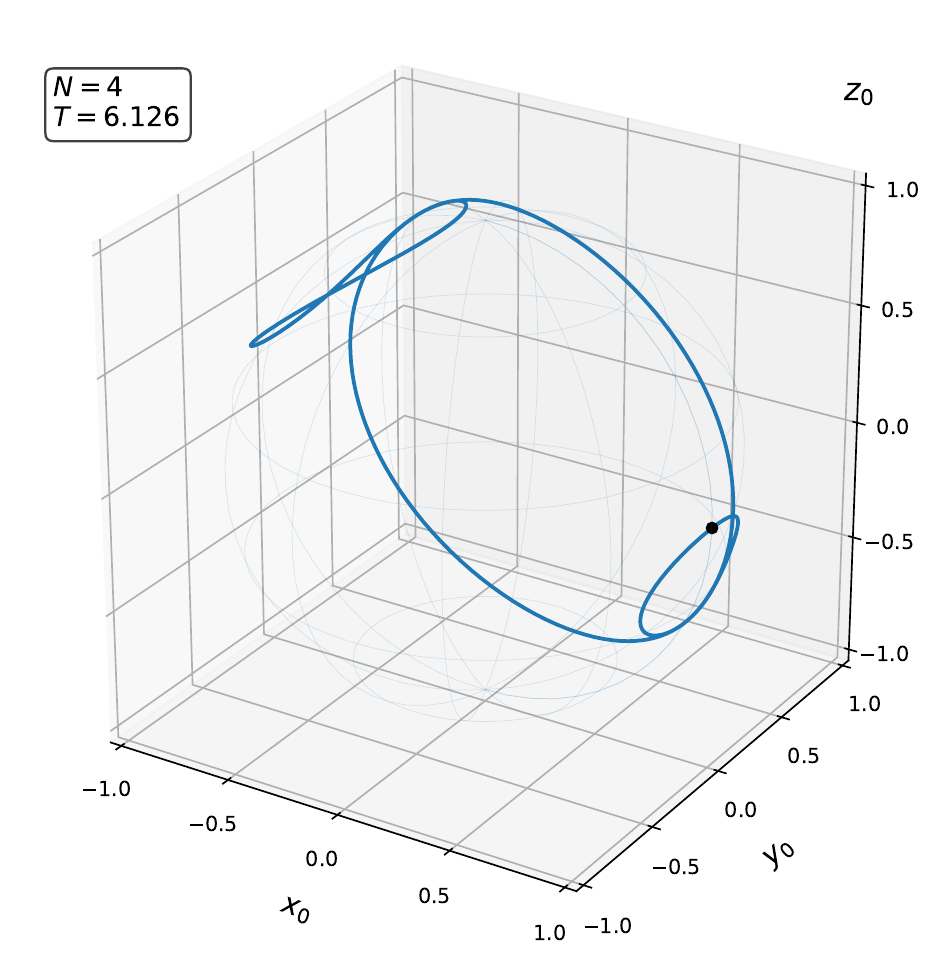}
    \end{minipage}
    \hfill
    \begin{minipage}[t]{0.32\textwidth}
        \centering
        \includegraphics[width=\textwidth]{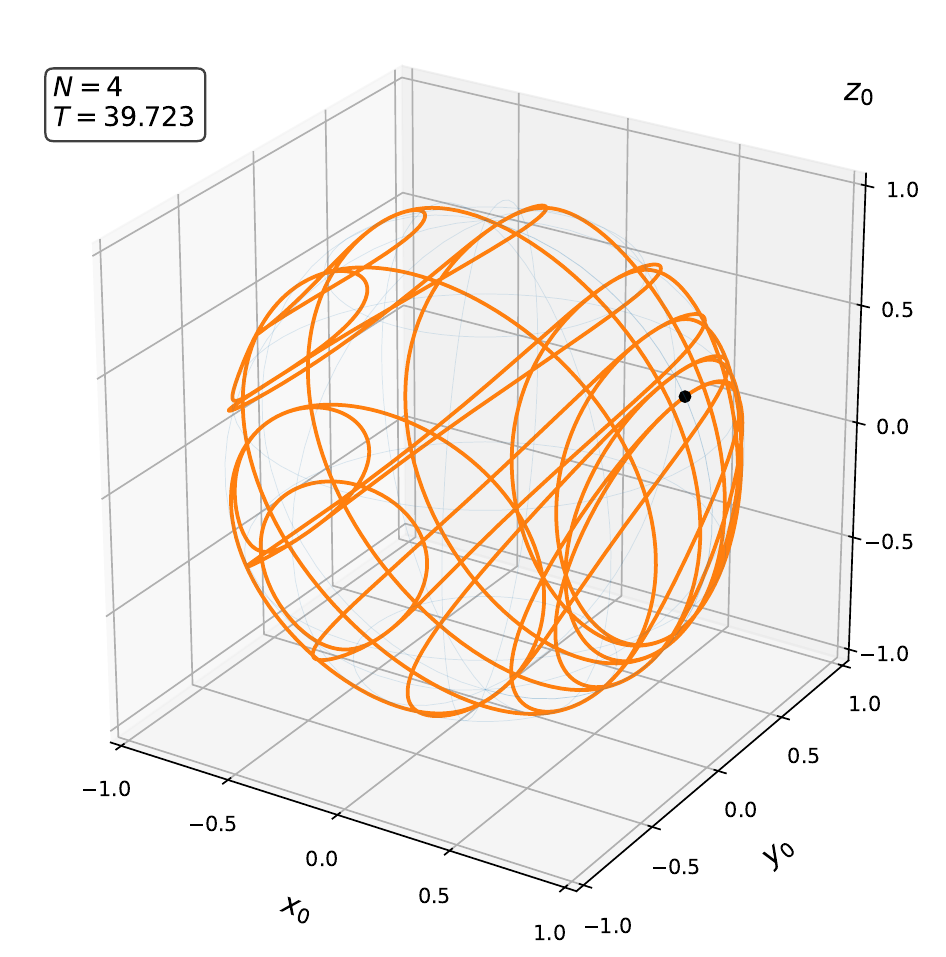}
    \end{minipage}
    \hfill
    \begin{minipage}[t]{0.32\textwidth}
        \centering
        \includegraphics[width=\textwidth]{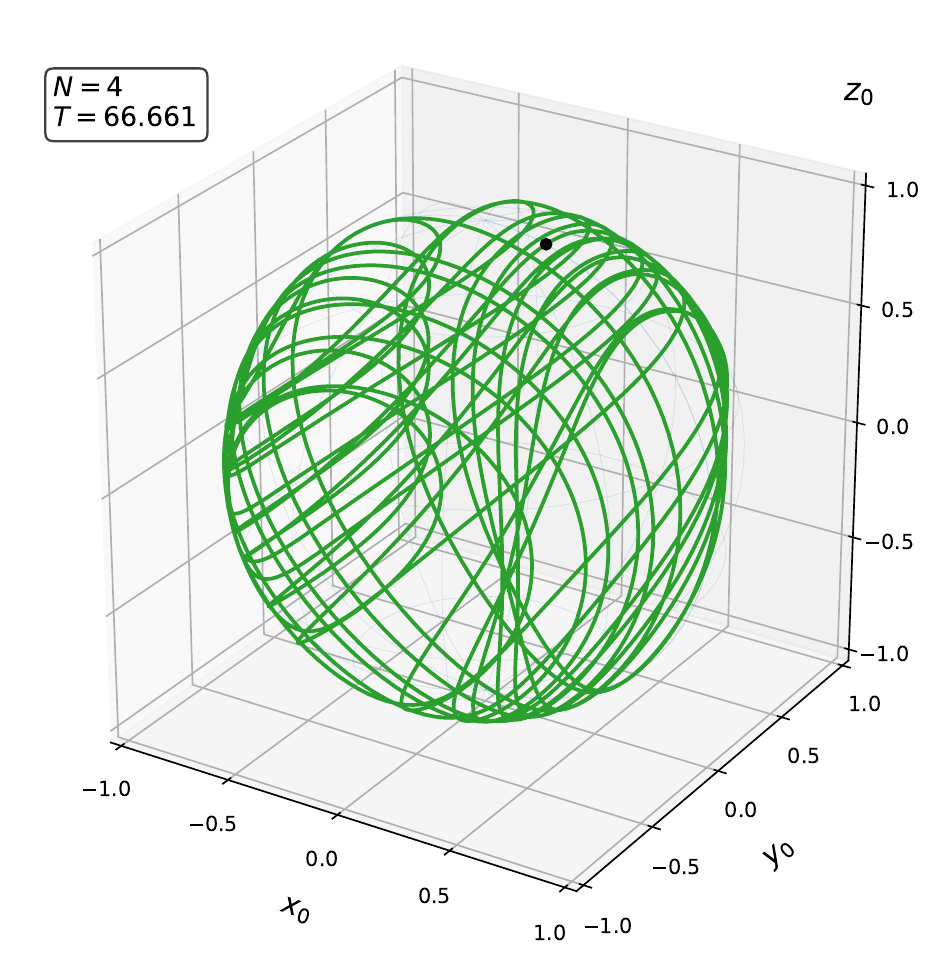}
    \end{minipage}

    \caption{\justifying
    Three representative unstable periodic orbits for $N=4$, shown through
    the Bloch trajectory of the central TLS in the coordinates
    $(x_0,y_0,z_0)$.  The value $T$ indicated in each panel is the period of the
    complete dynamical state, i.e., the time required for the full orbit to
    return to its initial condition.
    }
    \label{fig:representative_upos}
\end{figure*}

Figure~\ref{fig:representative_upos} shows three representative UPOs with
different periods and instability values. The figure illustrates the range
of recurrent motions found in the final catalogue, while the periodicity of
each orbit is validated in the full phase space.

% ------------------------------------------------------------
\subsection{Local instability of the periodic orbits}
\label{subsec:upo_stability}
% ------------------------------------------------------------

Finding a periodic trajectory does not yet characterize its local stability.
The selected orbits are therefore analyzed using Floquet theory
\cite{CvitanovicChaosBook2020,PilatowskyCameo2021}. The basic idea is to determine how an infinitesimal
perturbation of a point on the periodic orbit evolves after one complete
orbital period.

Let
\begin{equation}
    \bm{R}(0)
\end{equation}
be a point on a UPO and let
\begin{equation}
    \bm{R}(0)+\delta\bm{R}(0)
\end{equation}
be a nearby initial condition. After one complete orbital period $T$, the
UPO returns to its initial state,
\begin{equation}
    \bm{R}(T)=\bm{R}(0),
\end{equation}
whereas the small perturbation has evolved to
$\delta\bm{R}(T)$. Within the linear approximation, the relation between
the initial and final perturbations can be written as
\begin{equation}
    \delta\bm{R}(T)
    =
    \mathcal{M}(T)\,
    \delta\bm{R}(0),
    \label{eq:monodromy}
\end{equation}
where $\mathcal{M}(T)$ is the monodromy matrix of the periodic orbit. This
matrix describes how infinitesimal changes in the initial state are
transformed after one complete period.

The eigenvalues of the monodromy matrix, $\mu_j$, are the Floquet
multipliers \cite{CvitanovicChaosBook2020,PilatowskyCameo2021}. Several multipliers are obtained because
the initial state can be perturbed along different independent directions.
A multiplier satisfying
\begin{equation}
    |\mu_j|>1
\end{equation}
corresponds to a direction in which the perturbation grows after one period
and therefore indicates local instability. Conversely, $|\mu_j|<1$
corresponds to a contracting direction, while $|\mu_j|=1$ describes a
neutral direction.

An additional point must be considered for the homogeneous model studied
here. The environmental TLSs possess conserved geometrical quantities given
by the pairwise scalar products
\begin{equation}
    G_{ij}
    =
    \bm{r}_i\cdot\bm{r}_j,
    \qquad
    i,j=1,\ldots,N,
    \label{eq:gram_invariant_floquet}
\end{equation}
which are the elements of the environmental Gram matrix
\cite{HornJohnson2013}. Because all environmental TLSs experience the same
instantaneous effective field in the homogeneous model, these quantities
remain constant during the evolution. Their conservation and the resulting
collective structure are derived in
Appendix~\ref{app:collective_structure}.

This conservation must be considered when defining the perturbation
$\delta\bm{R}(0)$. A nearby state
\begin{equation}
    \bm{R}(0)+\delta\bm{R}(0)
\end{equation}
is a valid initial condition. However, if the environmental Bloch vectors are
perturbed independently, their relative angles generally change. Since $G_{ij}=\bm r_i\cdot\bm r_j$,
such a perturbation will in general modify the values of $G_{ij}$. The
perturbed state would then belong to a different invariant sector from the
original UPO. To construct a fixed-sector perturbation, $\delta\bm R(0)$ must
therefore be restricted so that all $G_{ij}$ remain unchanged.

For this reason, the Floquet calculation considers perturbations that
preserve the Gram invariants of the UPO. This is referred to as the fixed-sector Floquet analysis. It measures the local stability of the periodic orbit with respect to nearby trajectories belonging to the same invariant sector.

Some Floquet multipliers correspond to neutral directions and are not used
to quantify the instability. The largest modulus among the remaining fixed-sector multipliers is denoted by
\begin{equation}
    \rho_{\mathrm{sec}}
    =
    \max_{\mathrm{nontrivial}\,j}
    |\mu_j|.
\end{equation}
The corresponding local instability exponent is
\begin{equation}
    \lambda_{\mathrm{sec}}
    =
    \frac{\ln\rho_{\mathrm{sec}}}{T}.
    \label{eq:upo_lyapunov}
\end{equation}
A larger value of $\lambda_{\mathrm{sec}}$ therefore means that a small
perturbation separates more rapidly from the periodic orbit.

Because the selected UPOs have different periods, comparing
$\lambda_{\mathrm{sec}}$ alone does not provide the complete picture. The instability is therefore normalized by the orbital angular frequency,
\begin{equation}
    \omega
    =
    \frac{2\pi}{T},
\end{equation}
and define the dimensionless instability parameter
\begin{equation}
    \chi_{\mathrm{UPO}}
    =
    \frac{\lambda_{\mathrm{sec}}}{\omega}
    =
    \frac{\ln\rho_{\mathrm{sec}}}{2\pi}.
    \label{eq:lambda_omega}
\end{equation}

This quantity compares the local instability rate of the periodic orbit with its characteristic orbital frequency. Since the selected periodic orbits are already identified as unstable through their Floquet multipliers, $\chi_{\mathrm{UPO}}$ is not used to determine whether an orbit is stable or unstable. Instead, it provides a dimensionless measure of the strength of that instability relative to the orbital timescale. Smaller values indicate weaker divergence of nearby trajectories over one orbital cycle, whereas larger values correspond to stronger relative instability. In UPO-based studies, the condition
\begin{equation}
    \chi_{\mathrm{UPO}}
    =
    \frac{\lambda_{\mathrm{sec}}}{\omega}
    <1
\end{equation}
is commonly associated with conditions favorable for the persistence of periodic-orbit signatures \cite{Kaplan1999,Hummel2023,Pizzi2025}. Accordingly, $\chi_{\mathrm{UPO}}$ is used here to compare the relative instability of the identified UPOs on a common dimensionless scale.

\begin{figure}[h]
    \centering    \includegraphics[width=0.50\columnwidth]{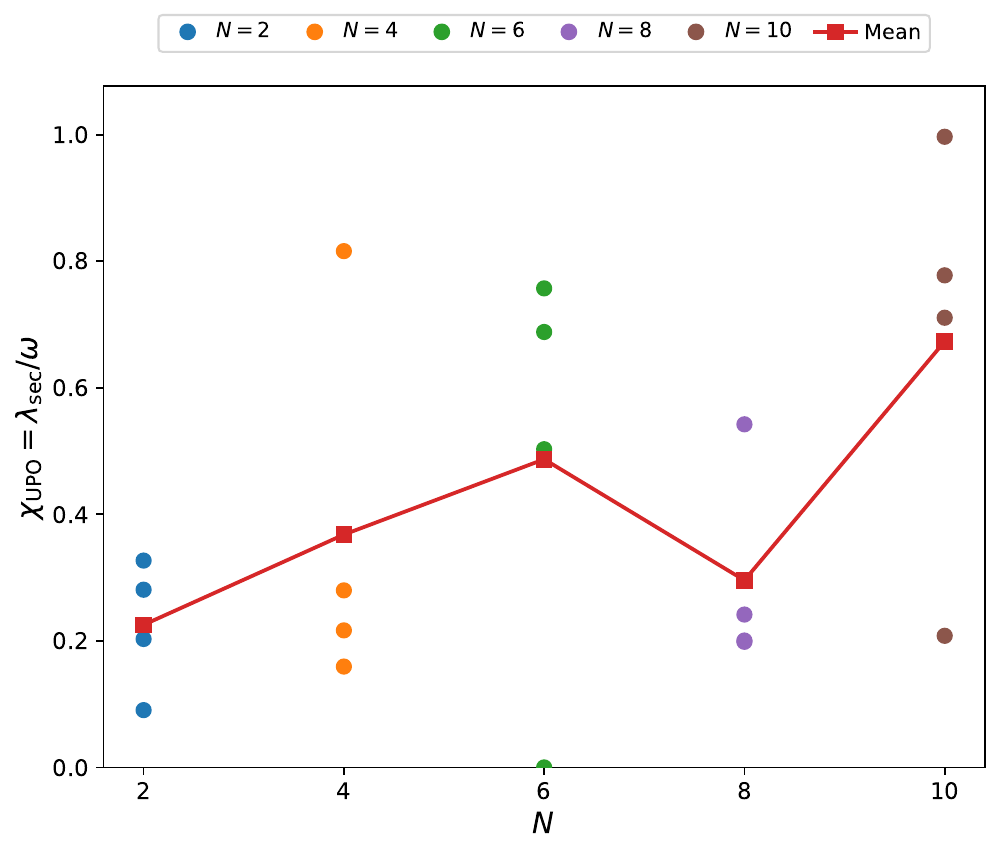}
    \caption{\justifying
    Local instability of the final collective UPO representatives as a
    function of the environment size $N$. Colored circles represent the four
    individual UPO families obtained for each $N$, with the colors identified
    in the legend. Red squares connected by a line show the mean value for
    each environment size. The instability is quantified by the dimensionless
    parameter
$\chi_{\mathrm{UPO}}=\lambda_{\mathrm{sec}}/\omega$.
    }
    \label{fig:upo_instability}
\end{figure}

As summarized in Fig.~\ref{fig:upo_instability}, the selected periodic
orbits display a broad range of instability values. The mean
$\chi_{\mathrm{UPO}}$ does not vary monotonically with $N$, so no simple
dependence of the average UPO instability on the number of environmental
TLSs can be established from the present set of periodic orbits.

Both the chaoticity of the surrounding dynamics, through the FTLE, and the local instability of the periodic orbits, through the Floquet analysis, have therefore been characterized. The next step is to determine whether these
periodic structures also organize the dynamics of nearby chaotic
trajectories. In the following section, trajectories initialized close to each UPO are compared with corresponding reference trajectories to test whether they retain a preferential spatial association with the periodic path and temporal recurrences related to its fundamental period.
% ============================================================
\section{UPO-associated dynamical localization}
\label{sec:dynamical_localization}
% ============================================================

The previous section characterized the chaotic background of the system and
the local instability of the UPOs embedded within it. A different question is now considered in order to extend this analysis: can these periodic orbits also influence the evolution of nearby chaotic trajectories? In particular, it is investigated whether trajectories initialized close to a UPO retain measurable spatial or temporal properties associated with that orbit after evolving away from their initial conditions.

The term \emph{UPO memory} is used to describe this behavior. A trajectory
displays UPO memory when, after the immediate effect of its initial proximity
to the orbit has been excluded, its subsequent dynamics retains measurable
properties associated with that particular periodic orbit. Such behavior
would indicate that the UPO contributes to organizing the surrounding
chaotic dynamics.

This question is closely related to the periodic-orbit picture of quantum
scarring. In conventional scar studies, the influence of an unstable
periodic orbit is investigated through quantities such as enhanced
phase-space weight around the orbit, recurrent returns associated with its
period, and frequency components related to the orbital motion
\cite{Heller1984,KaplanHeller1999,PilatowskyCameo2021,Evrard2024,Pizzi2025}.
Here, analogous dynamical quantities adapted to the nonlinear system--environment model are used.

For each of the 20 UPO families identified in Sec.~\ref{sec:upos}, two
ensembles of trajectories are constructed. The first, referred to as the
\emph{near-UPO ensemble}, consists of trajectories initialized in the
immediate vicinity of the periodic orbit. The second, referred to as the
\emph{matched-reference ensemble}, consists of trajectories initialized away
from the UPO while belonging to the same fixed-energy invariant sector.

Consequently,
\begin{equation}
    H_{\mathrm{near}}
    =
    H_{\mathrm{ref}}
    =
    H_{\gamma},
\end{equation}
and
\begin{equation}
    G_{ij}^{\mathrm{near}}
    =
    G_{ij}^{\mathrm{ref}}
    =
    G_{ij}^{\gamma}.
    \label{eq:matched_invariants}
\end{equation}
The relevant difference between the two ensembles is therefore their initial
relation to the UPO rather than their energy or invariant sector. The
specific procedure used to construct these initial conditions is described
below, after introducing Eq. \eqref{eq:product_state_overlap}.

For every UPO,
\begin{equation}
    N_{\mathrm{near}}=24
    \qquad\mathrm{and}\qquad
    N_{\mathrm{ref}}=48
\end{equation}
independent trajectories are propagated for a total time of $20T$, where $T$
is the fundamental period of the corresponding UPO. Once the initial
conditions have been generated, both ensembles evolve freely according to the
same nonlinear equations of motion. In particular, the near-UPO trajectories
are not forced to remain close to the periodic orbit at any later time.

To compare complete system--environment configurations, the product-state
overlap is defined. Let
\begin{equation}
    \bm{R}
    =
    (\bm{r}_0,\bm{r}_1,\ldots,\bm{r}_N)
\end{equation}
and
\begin{equation}
    \bm{R}'
    =
    (\bm{r}'_0,\bm{r}'_1,\ldots,\bm{r}'_N)
\end{equation}
denote two different complete system--environment configurations. Their
overlap is defined as
\begin{equation}
    q(\bm{R},\bm{R}')
    =
    \prod_{k=0}^{N}
    \frac{
        1+\bm{r}_k\cdot\bm{r}'_k
    }{2},
    \label{eq:product_state_overlap}
\end{equation}
where $k=0$ corresponds to the central TLS and $k=1,\ldots,N$ to the
environmental TLSs \cite{Arecchi1972,Pizzi2025}. This quantity measures the similarity between two
complete system--environment configurations. In particular, $q=1$
corresponds to identical configurations, whereas decreasing values of $q$
indicate progressively smaller similarity.

The near-UPO and matched-reference initial conditions can now be specified
more precisely. For the near-UPO ensemble, a random point along the periodic
orbit is selected and slightly perturbed. The perturbation corresponds to a
characteristic angular displacement of $10^{-4}\,\mathrm{rad}$ of the Bloch
vectors and is constructed so that the total energy and Gram invariants of
the UPO remain unchanged. 

For the matched-reference ensemble, initial conditions are sampled from the
same fixed-energy invariant sector but are required to remain sufficiently
far from the complete UPO. Using the product-state overlap defined in
Eq.~\eqref{eq:product_state_overlap}, a candidate reference state is accepted
only if its maximum overlap with the periodic orbit does not exceed $0.20$.
The reference trajectories therefore share the same energy and Gram
invariants as the near-UPO trajectories but do not initially lie in the
immediate neighborhood of the UPO.

Several complementary quantities are used to determine whether trajectories initialized close to a UPO retain information about its spatial path and characteristic period. Phase-space localization and recurrence-based
quantities are commonly used to characterize periodic-orbit signatures
\cite{KaplanHeller1999,PilatowskyCameo2021,PilatowskyCameo2022,
Evrard2024,Pizzi2025}. Other quantities, including spectral measures of the
orbital frequency, can provide additional information, but the analysis here
focuses on the spatial localization and recurrence observables introduced
below.
\vspace{1cm}
\begin{enumerate}

\item \textbf{Full-window tube localization.}

The first quantity measures the association between a trajectory and the
complete periodic path rather than with one particular point of the
UPO. A given periodic orbit is labeled by $\gamma$ and represented using
$N_{\phi}=360$ uniformly distributed points,
\begin{equation}
    \left\{
    \bm{R}_{\gamma}(\theta_j)
    \right\}_{j=1}^{N_{\phi}},
\end{equation}
where $\theta_j$ specifies the position along the periodic orbit.

For any trajectory $\bm{R}(t)$, either from the near-UPO or reference ensemble, the quantity
\begin{equation}
    Q_{\gamma}(t)
    =
    \frac{1}{N_{\phi}}
    \sum_{j=1}^{N_{\phi}}
    q\left(
        \bm{R}(t),
        \bm{R}_{\gamma}(\theta_j)
    \right)
    \label{eq:tube_projection}
\end{equation}
is defined.
Each term in the sum compares the state of the trajectory at time $t$ with
one point of the UPO. Averaging over the 360 points provides a single measure
of how strongly that trajectory is associated with the complete periodic
path. A larger value of $Q_{\gamma}(t)$ therefore means that the trajectory
at time $t$ has a stronger association with UPO $\gamma$.

This provides an orbit-specific measure analogous to the enhanced
phase-space weight used in periodic-orbit scar analyses
\cite{KaplanHeller1999,PilatowskyCameo2021,Pizzi2025}..

For a time interval $[aT,bT]$, the time-averaged value
\begin{equation}
    \overline{Q}_{\gamma}^{[a,b]}
    =
    \frac{1}{(b-a)T}
    \int_{aT}^{bT}
    Q_{\gamma}(t)\,dt
    \label{eq:tube_time_average}
\end{equation}
is defined, and the two ensembles are compared through
\begin{equation}
    \mathcal{P}_{\mathrm{tube}}^{[a,b]}
    =
    \frac{
        \left\langle
        \overline{Q}_{\gamma}^{[a,b]}
        \right\rangle_{\mathrm{near}}
    }{
        \left\langle
        \overline{Q}_{\gamma}^{[a,b]}
        \right\rangle_{\mathrm{ref}}
    }.
    \label{eq:tube_enhancement}
\end{equation}
Thus,
\begin{equation}
    \mathcal{P}_{\mathrm{tube}}^{[a,b]}>1
\end{equation}
indicates that trajectories initialized near the UPO remain more strongly
associated with its periodic path than the matched-reference trajectories.

The main interval considered here is $2T$--$20T$. The interval from $0$ to
$2T$ is excluded because the near-UPO ensemble is deliberately initialized
very close to the periodic orbit. A large value of $Q_{\gamma}(t)$ at very
early times could therefore simply reflect how the initial conditions were
constructed. By starting the quantitative analysis at $2T$, it is tested whether the stronger association with the UPO remains visible after the trajectories have evolved freely for two complete orbital periods.

\item \textbf{Late-time tube localization.}

To determine whether this spatial association remains detectable much later in the evolution, the same analysis is repeated using only the final interval $16T$--$20T$. The corresponding quantity is
\begin{equation}
    \mathcal{P}_{\mathrm{tube}}^{[16,20]}.
\end{equation}
This provides a more restrictive test of persistent UPO-associated
localization because it measures the difference between the two ensembles
long after their initialization.

\item \textbf{Return fidelity.}

Tube localization measures association with the complete periodic path. It is next examined whether a trajectory also returns close to its own initial state.
This is quantified through the return fidelity
\begin{equation}
    F(t)
    =
    q\left(
        \bm{R}(t),
        \bm{R}(0)
    \right)
    =
    \prod_{k=0}^{N}
    \frac{
        1+\bm{r}_k(0)\cdot\bm{r}_k(t)
    }{2}.
    \label{eq:return_fidelity}
\end{equation}
By construction, $F(0)=1$. Values close to unity at later times indicate
that the complete system--environment configuration has returned close to
its initial state. Return fidelities or survival probabilities around
integer multiples of a UPO period are standard dynamical signatures of
periodic-orbit memory
\cite{PilatowskyCameo2021,Evrard2024}.

Because the exact UPO returns after every integer multiple of its period, the first five expected recurrences are analyzed,
\begin{equation}
    T,\;2T,\;\ldots,\;5T.
\end{equation}
For each recurrence $n$, $F_n^{\mathrm{max}}$ is defined as the largest value of the return fidelity within a small interval around $nT$,
\begin{equation}
    F_n^{\mathrm{max}}
    =
    \max_{|t-nT|\leq \eta T}
    F(t),
    \qquad
    \eta=0.12.
    \label{eq:return_peak}
\end{equation}
This finite time window is used because a trajectory that is not exactly
periodic is not expected to reach its maximum return fidelity precisely at
an integer multiple of the UPO period. The maximum can instead occur slightly
before or after $nT$, while still remaining dynamically associated with the
recurrence imposed by the periodic orbit.

The average amplitude of the first five returns is then
\begin{equation}
    A_{1:5}
    =
    \frac{1}{5}
    \sum_{n=1}^{5}
    F_n^{\mathrm{max}}.
    \label{eq:return_amplitude}
\end{equation}
The relative enhancement between the two ensembles is defined as
\begin{equation}
    \mathcal{P}_{F}^{1:5}
    =
    \frac{
        \left\langle A_{1:5}\right\rangle_{\mathrm{near}}
    }{
        \left\langle A_{1:5}\right\rangle_{\mathrm{ref}}
    }.
    \label{eq:return_enhancement}
\end{equation}
Therefore,
\begin{equation}
    \mathcal{P}_{F}^{1:5}>1
\end{equation}
means that trajectories initialized close to the UPO exhibit stronger
returns toward their initial configurations than dynamically comparable
reference trajectories. This provides a temporal signature that the
periodic orbit is influencing the surrounding chaotic dynamics.

\item \textbf{Return timing.}

The amplitude of a return does not determine how accurately that return
follows the period of the UPO. The positions of the first five return fidelity maxima are therefore also compared with the recurrence times predicted by the periodic orbit.

If $t_n^{*}$ denotes the time at which $F_n^{\mathrm{max}}$ occurs, the normalized timing error is defined as
\begin{equation}
    D_{1:5}
    =
    \frac{1}{5}
    \sum_{n=1}^{5}
    \frac{
        |t_n^{*}-nT|
    }{T}.
    \label{eq:return_timing}
\end{equation}
A smaller value of $D_{1:5}$ means that the observed recurrences occur more
closely to the times
\begin{equation}
    T,\;2T,\;\ldots,\;5T
\end{equation}
defined by the periodic orbit. UPO-associated temporal memory therefore
corresponds to
\begin{equation}
    \left\langle D_{1:5}\right\rangle_{\mathrm{near}}
    <
    \left\langle D_{1:5}\right\rangle_{\mathrm{ref}}.
\end{equation}
This quantity provides a direct test of whether the fundamental period of
the UPO continues to organize the timing of nearby chaotic trajectories
\cite{PilatowskyCameo2021,Evrard2024}.

\end{enumerate}

For each UPO family, the near-UPO and matched-reference ensembles contain
several independent trajectories. Comparing only their mean values is not
enough, because one ensemble may have a larger mean while the individual
values within both ensembles are widely dispersed. A Mann--Whitney test \cite{MannWhitney1947} is therefore used to assess whether the two ensembles differ consistently.

A UPO family is classified as \textit{significant} for a given diagnostic
when the corresponding Mann--Whitney test gives
\begin{equation}
    p_{\mathrm{MW}}<0.05,
\end{equation}
where $p_{\mathrm{MW}}$ is the parameter provided by the test. In practical
terms, this indicates that the criterion associated with that diagnostic is
statistically supported by the comparison between the near-UPO and
matched-reference ensembles. In summary, \textit{significant} means that the near-UPO ensemble shows a statistically supported difference in the expected direction: higher full-window and late-time tube localization, higher return fidelity, or lower return timing than the matched-reference ensemble.

The main results are summarized in Table~\ref{tab:localization_summary}.

\begin{table}[H]
    \centering
    \small

    \begin{tabular}{lcc}
        \hline
        \textbf{Diagnostic} &
        \textbf{UPO-memory criterion} &
        \textbf{Significant families} \\
        \hline

        Tube localization $2T$--$20T$ &
        $\mathcal{P}_{\mathrm{tube}}^{[2,20]}>1$ &
        18/20 \\

        Tube localization $16T$--$20T$ &
        $\mathcal{P}_{\mathrm{tube}}^{[16,20]}>1$ &
        14/20 \\

        Return fidelity $1T$--$5T$ &
        $\mathcal{P}_{F}^{1:5}>1$ &
        19/20 \\

        Return timing $1T$--$5T$ &
        $D_{1:5}^{\mathrm{near}}<D_{1:5}^{\mathrm{ref}}$ &
        20/20 \\

        \hline
    \end{tabular}

    \caption{\justifying
    Summary of the main diagnostics used to quantify UPO-associated
    dynamical localization. The last column gives the number of UPO families
    showing a statistically significant difference between the near-UPO and
    matched-reference ensembles.
    }
    \label{tab:localization_summary}
\end{table}

As shown in Table~\ref{tab:localization_summary}, 18 of the 20 UPO families
show an enhanced occupation of the complete periodic path over the interval
$2T$--$20T$. This effect remains statistically detectable during the final
$16T$--$20T$ interval for 14 families, showing that the spatial association
with the UPO often persists long after initialization.

The return fidelity analysis provides a complementary temporal result.
Enhanced return amplitudes over the first five periods are found for 19 of
the 20 UPO families. Furthermore, the recurrence timing remains associated
with the UPO period for all 20 families: the return maxima of the near-UPO
trajectories occur significantly closer to $T,2T,\ldots,5T$ than those of
the matched-reference trajectories.

The results therefore indicate a specific form of dynamical organization:
trajectories initialized near a UPO retain both a preferential association
with its periodic path and recurrent dynamics related to its fundamental
period.

\begin{figure}[h]
    \centering
    \includegraphics[width=\linewidth]{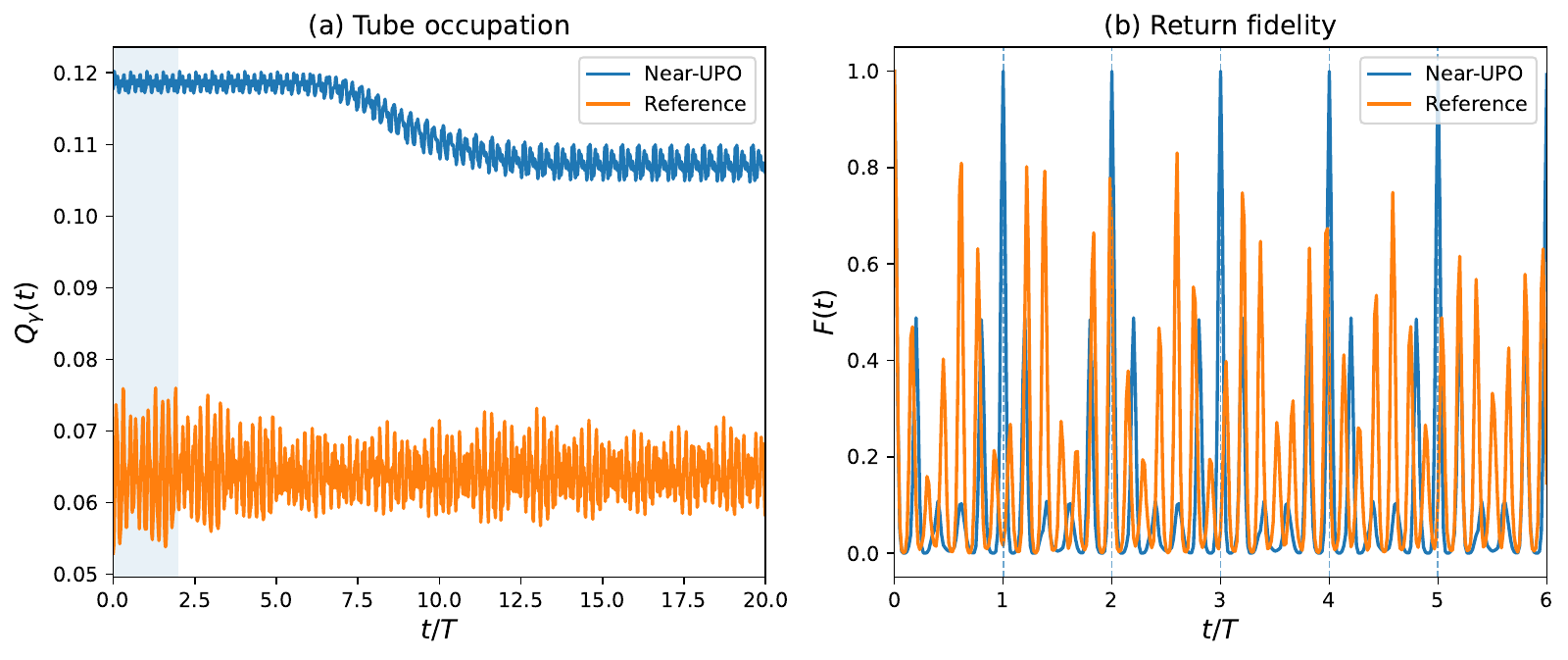}
    \caption{\justifying
    Representative visualization of UPO-associated dynamical memory for an
    $N=4$ periodic orbit with $T=27.8043$ and
    $\chi_{\mathrm{UPO}}=\lambda_{\mathrm{sec}}/\omega=0.2445$.
    (a) Phase-averaged tube occupation $Q_{\gamma}(t)$ for the near-UPO and
    matched-reference ensembles. The shaded interval
    $0\leq t/T\leq2$ indicates the initial interval excluded from the
    quantitative tube-localization analysis.
    (b) Return fidelity $F(t)$ for the same two ensembles. The vertical
    dashed lines indicate the recurrence times $t=nT$, with
    $n=1,\ldots,5$.
    }
    \label{fig:upo_memory}
\end{figure}

To illustrate these quantities directly, Fig.~\ref{fig:upo_memory} shows one
representative $N=4$ UPO. In panel~(a), the near-UPO trajectories retain a
larger tube occupation than the matched-reference trajectories after the
initial $0$--$2T$ interval excluded from the quantitative analysis. In
panel~(b), their return fidelity maxima remain associated with the recurrence
times $t=nT$. The figure therefore provides a representative example of the
spatial and temporal effects quantified statistically across the complete
set of UPO families.

Taken together, these results show that the periodic orbits identified in
Sec.~\ref{sec:upos} leave statistically reproducible spatial and temporal
signatures in nearby chaotic trajectories. In most cases, trajectories
initialized close to a UPO retain an enhanced association with its complete
periodic path, while their recurrences remain related to the fundamental
period of the orbit.

This behavior is referred to as \emph{UPO-associated dynamical localization}.
The term is used here to describe persistent orbit-specific spatial and
temporal memory in the nonlinear dynamics, rather than conventional
eigenstate scarring of a fixed linear Hamiltonian.

Having established this phenomenon at the reference energy density, the
following section examines how these signatures change when the energy is varied.

% ============================================================
\section{Energy dependence of chaos and UPO-associated localization}
\label{sec:energy_dependence}
% ============================================================

The previous section established that unstable periodic orbits leave
measurable spatial and temporal signatures in nearby chaotic trajectories at
the reference energy density. It is now examined whether these signatures change when the energy of the system is varied.

To study the effect of energy independently of the environment size, the number of environmental TLSs is fixed at
\begin{equation}
    N=10.
\end{equation}
Five energy shells centered on the reference energy density $e=0.8371$ are considered, including two lower- and two higher-energy cases. The selection of these energies from the accessible energy interval is described in Appendix~\ref{app:energy_ranges}.The corresponding energy densities are
\begin{equation}
    e
    =
    0.2152,\;
    0.5262,\;
    0.8371,\;
    1.1481,\;
    1.4590.
    \label{eq:energy_scan_density}
\end{equation}

At each energy density, four independent UPOs are selected and the same
near-UPO and matched-reference analysis introduced in
Sec.~\ref{sec:dynamical_localization} is repeated. The objective is not to
introduce new localization measures, but to determine how the previously
defined dynamical quantities change with energy. The results are summarized
in Table~\ref{tab:energy_scan_summary}. For each energy density, the table
reports the mean values obtained from the four selected UPOs, allowing the
evolution of chaoticity, local UPO instability, and UPO-associated
localization to be compared over the energy scan.

\begin{table}[h]
\centering
\small

\begin{tabular}{cccccc}
    \hline
    $e$ &
    $\langle\widetilde{\lambda}_{\mathrm{FT}}^{(\mathrm{sec})}\rangle$ &
    $\langle\chi_{\mathrm{UPO}}\rangle$ &
    $\langle\mathcal{P}_{\mathrm{tube}}^{[2,20]}\rangle$ &
    $\langle\mathcal{P}_{F}^{1:5}\rangle$ &
    Timing \\
    \hline

    0.2152 & 0.0517 & 0.3796 & 1.668 & 3.169 & 4/4 \\
    0.5262 & 0.1012 & 0.4975 & 1.695 & 3.151 & 4/4 \\
    0.8371 & 0.2092 & 0.6733 & 1.065 & 1.779 & 4/4 \\
    1.1481 & 0.3240 & 0.8625 & 1.033 & 1.523 & 4/4 \\
    1.4590 & 0.4625 & 0.5619 & 1.045 & 1.777 & 4/4 \\

    \hline
\end{tabular}

\caption{\justifying
Energy dependence of the main dynamical and localization diagnostics for
$N=10$. Four independent UPOs are considered at each energy density.
$\widetilde{\lambda}_{\mathrm{FT}}^{(\mathrm{sec})}$ denotes the median
finite-time Lyapunov exponent evaluated within the invariant sector, while
$\chi_{\mathrm{UPO}}=\lambda_{\mathrm{sec}}/\omega$ quantifies the
dimensionless local instability of the periodic orbit. The angle brackets
denote the mean over the four UPOs selected at each energy.
$\mathcal{P}_{\mathrm{tube}}^{[2,20]}$ is the tube-localization enhancement
over the interval $2T$--$20T$, and
$\mathcal{P}_{F}^{1:5}$ is the enhancement of the first five
return fidelity peaks. The final column gives the number of UPO families,
out of four, satisfying the statistically significant recurrence-timing
criterion introduced in Sec.~\ref{sec:dynamical_localization}.
}
\label{tab:energy_scan_summary}

\end{table}

As can be seen in Table~\ref{tab:energy_scan_summary}, the chaoticity of the
system shows a clear dependence on energy. For each energy density, the
sector FTLE is calculated for the four selected UPO families and their median
values are averaged. The resulting
$\langle\widetilde{\lambda}_{\mathrm{FT}}^{(\mathrm{sec})}\rangle$ increases
monotonically from $0.0517$ at the lowest energy to $0.4625$ at the highest
one. The dynamics therefore becomes progressively more chaotic over the
energy interval considered.

The remaining quantities do not vary monotonically with energy. The mean
local UPO instability, $\langle\chi_{\mathrm{UPO}}\rangle$, increases from
$0.3796$ to $0.8625$ over the first four energy densities, before decreasing
to $0.5619$ at the highest energy. The selected UPOs therefore tend to become
more unstable as the energy is increased, although this tendency does not
extend monotonically over the complete scan.

The two localization measures display the opposite general behavior. The
mean return fidelity enhancement
$\langle\mathcal{P}_{F}^{1:5}\rangle$ decreases from values above $3$ at the
two lowest energies to values between approximately $1.5$ and $1.8$ at the
three higher energies. Similarly,
$\langle\mathcal{P}_{\mathrm{tube}}^{[2,20]}\rangle$ takes values of $1.668$
and $1.695$ at the two lowest energies and remains much closer to unity at the
three higher energies. Neither dependence is strictly monotonic, since both
quantities increase again at the highest energy, but the overall trend is a
reduction of the UPO-associated localization strength as the energy is
increased.

\begin{figure}[h]
    \centering
    \includegraphics[width=0.8\linewidth]{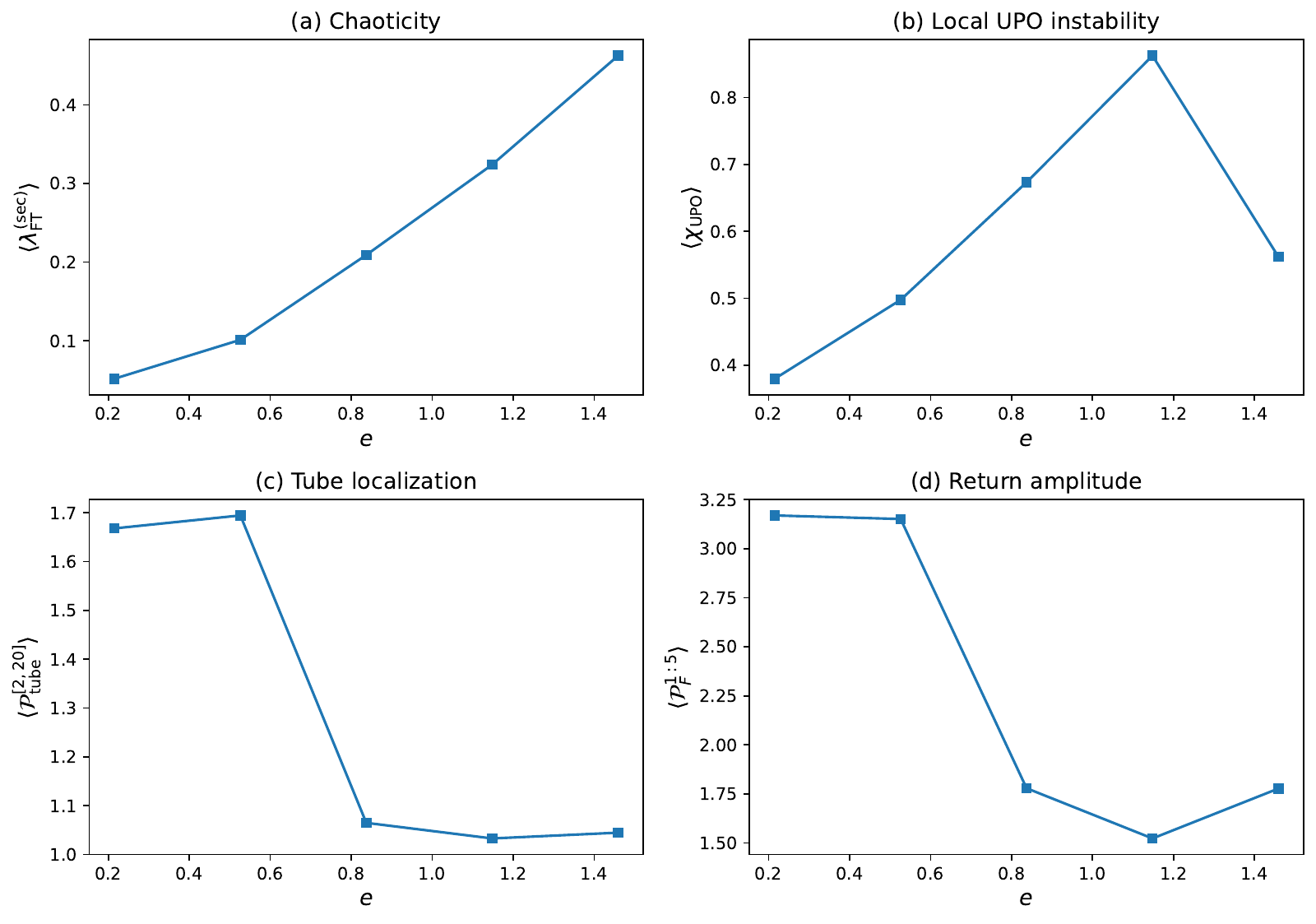}
    \caption{\justifying
    Energy dependence of chaoticity, local UPO instability, and
    UPO-associated dynamical localization for $N=10$.
    (a) Mean invariant-sector finite-time Lyapunov exponent
    $\langle\widetilde{\lambda}_{\mathrm{FT}}^{(\mathrm{sec})}\rangle$
    as a function of the energy density $e$.
    (b) Mean dimensionless local UPO instability
    $\langle\chi_{\mathrm{UPO}}\rangle$, with
    $\chi_{\mathrm{UPO}}=\lambda_{\mathrm{sec}}/\omega$.
    (c) Mean tube-localization enhancement
    $\langle\mathcal{P}_{\mathrm{tube}}^{[2,20]}\rangle$.
    (d) Mean enhancement of the first five return fidelity peaks,
    $\langle\mathcal{P}_{F}^{1:5}\rangle$.
    In all panels, each point represents the mean over the four collectively
    inequivalent UPO families obtained at the corresponding energy density.
    }
    \label{fig:energy_dependence}
\end{figure}

These trends are shown directly in Fig.~\ref{fig:energy_dependence}, which
compares the energy dependence of the chaoticity, the local Floquet
instability of the UPOs, and the two localization enhancements.

The recurrence timing shows a different behavior. At every energy considered,
all four UPO families satisfy the timing criterion introduced in
Sec.~\ref{sec:dynamical_localization}. Thus, for all 20 periodic orbits in
the energy scan, the first five return fidelity maxima of the near-UPO
trajectories occur significantly closer to
\begin{equation}
    T,\;2T,\;\ldots,\;5T
\end{equation}
than those of the corresponding matched-reference trajectories.

The energy scan therefore shows that increasing energy is accompanied by a
strong increase in chaoticity and, in general, by weaker tube-localization
and return fidelity enhancements. The local Floquet instability also tends to
increase over most of the scan, although its dependence is not monotonic.
Despite these changes in magnitude, the recurrence timing remains
statistically significant for every UPO considered, showing that the
characteristic period of the UPO continues to influence nearby chaotic
trajectories throughout the energy interval studied.

% ============================================================
\section{Robustness to central bias and environmental inhomogeneity}
\label{sec:robustness}
% ============================================================

The previous sections established UPO-associated dynamical localization in
the homogeneous model and examined how its main signatures vary with the
energy density. It is now examined whether these signatures persist when the bias parameters of the model are changed.

This question is addressed in two steps. First, a finite bias is introduced in the central TLS while all environmental TLSs are kept identical. Second, different biases are also introduced among the environmental TLSs, thereby breaking the homogeneity of the environment.

For the parameter values considered throughout this work, the complete
Hamiltonian can be written as
\begin{equation}
    H
    =
    -2x_0+2\epsilon z_0
    +
    \sum_{i=1}^{N}
    \left(
        -2x_i+2\epsilon_i z_i
    \right)
    +
    z_0\sum_{i=1}^{N}z_i ,
    \label{eq:asymmetric_hamiltonian}
\end{equation}
where $\epsilon$ is the energetic bias of the central TLS and $\epsilon_i$
is the corresponding bias of the $i$th environmental TLS.

The homogeneous model considered in the previous sections corresponds to
\begin{equation}
    \epsilon=0,
    \qquad
    \epsilon_i=0.
    \label{eq:baseline_bias}
\end{equation}
Because all environmental TLSs have identical dynamical parameters, their
pairwise Gram quantities
\begin{equation}
    G_{ij}
    =
    \bm{r}_i\cdot\bm{r}_j
\end{equation}
remain constant during the evolution. As discussed previously, this
invariant structure must be respected when constructing perturbations for
the Floquet analysis and when generating dynamically comparable reference
trajectories.

In the first modified model, a bias is introduced only in the central TLS,
\begin{equation}
    \epsilon=1,
    \qquad
    \epsilon_i=0.
    \label{eq:central_bias_model}
\end{equation}
Although the central dynamics is modified, all environmental TLSs remain
identical and continue to experience the same instantaneous effective field.
Consequently, the Gram invariants remain conserved and the same collective
description used in the homogeneous reference model continues to apply.

In the second modified model, the central bias is retained and the
environment is made inhomogeneous by assigning a different bias to each
environmental TLS,
\begin{equation}
    \epsilon=1,
    \qquad
    \epsilon_i
    =
    \sigma_{\epsilon}\xi_i,
    \qquad
    \sigma_{\epsilon}=0.1,
    \label{eq:inhomogeneous_bias_model}
\end{equation}
where the deterministic set $\{\xi_i\}$ is chosen to have zero mean and unit
root-mean-square amplitude,
\begin{equation}
    \frac{1}{N}\sum_{i=1}^{N}\xi_i=0,
    \qquad
    \frac{1}{N}\sum_{i=1}^{N}\xi_i^2=1.
\end{equation}
The environmental TLSs therefore no longer have identical dynamical
parameters. In this case they do not, in general, experience the same
instantaneous effective field, and the pairwise quantities $G_{ij}$ are no
longer constants of motion.

For both modified models, the calculations are performed at the reference
energy density used in the main analysis,
\begin{equation}
    e
    =
    \frac{E}{N+1}
    =
    0.83711739,
\end{equation}
and for
\begin{equation}
    N=2,4,6,8,10.
\end{equation}
For each value of $N$, four independent UPO families are considered, giving
20 UPO families for each model. The same dynamical observables introduced in
Sec.~\ref{sec:dynamical_localization} are then evaluated.

One difference must be considered when constructing the reference
ensembles. In the two homogeneous models, the near-UPO and reference
trajectories are matched in both energy and Gram invariants. Once the
environmental homogeneity is broken, the Gram quantities are no longer
conserved, and the reference trajectories are therefore matched only in
energy. The comparison below focuses on whether the same dynamical
signatures remain statistically detectable in each model.

Following the same analysis used in the previous sections, the main results
are summarized in Table~\ref{tab:robustness_summary}.

\begin{table}[h]
\centering
\small

\begin{tabular}{lccc}
    \hline
    &
    Homogeneous &
    Central bias &
    Inhomogeneous environment \\
    \hline

    $\epsilon$
    & 0 & 1 & 1 \\

    $\epsilon_i$
    & 0 & 0 & $0.1\xi_i$ \\

    $\mathcal{P}_{\mathrm{tube}}^{[2,20]}$
    & 18/20 & 20/20 & 20/20 \\

    $\mathcal{P}_{\mathrm{tube}}^{[16,20]}$
    & 14/20 & 17/20 & 20/20 \\

    $\mathcal{P}_{F}^{1:5}$
    & 19/20 & 17/20 & 19/20 \\

    $D_{1:5}$
    & 20/20 & 20/20 & 20/20 \\

    \hline
\end{tabular}

\caption{\justifying
Comparison of the main UPO-associated dynamical signatures in the
homogeneous reference model, the centrally biased homogeneous model,
and the environmentally inhomogeneous model. Each value indicates the number of statistically significant UPO families out of the 20 analyzed.
}
\label{tab:robustness_summary}

\end{table}

\begin{figure}[H]
    \centering
    \includegraphics[width=\linewidth]{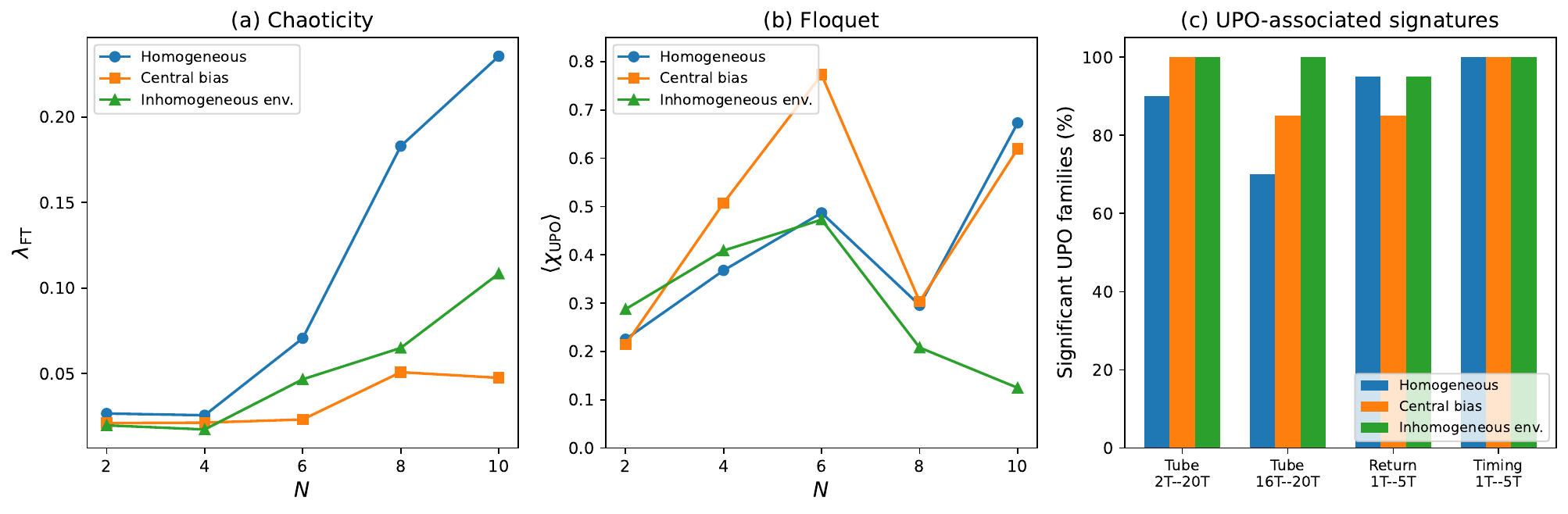}
    \caption{\justifying
    Robustness of the chaotic background, local UPO instability, and
    UPO-associated dynamical signatures under central bias and environmental
    inhomogeneity.
    (a) Median finite-time Lyapunov exponent
    $\widetilde{\lambda}_{\mathrm{FT}}$ of the same-energy trajectory ensemble
    as a function of the environment size $N$ for the homogeneous reference
    model, the centrally biased homogeneous model, and the environmentally
    inhomogeneous model.
    (b) Mean normalized local UPO instability of the four representative UPO
    families obtained at each $N$. For the homogeneous reference model and
    the centrally biased homogeneous model, the quantity shown is
    $\lambda_{\mathrm{sec}}/\omega$, evaluated within the corresponding
    invariant sector. For the environmentally inhomogeneous model, the
    corresponding full-space normalized Floquet exponent is shown.
    (c) Percentage of UPO families satisfying the statistical criterion
    (Mann--Whitney $p<0.05$) for the main dynamical
    diagnostics: tube localization over $2T$--$20T$, late-time tube
    localization over $16T$--$20T$, return fidelity enhancement over
    $1T$--$5T$, and recurrence timing over $1T$--$5T$.}
    \label{fig:robustness}
\end{figure}

The effect of the two modifications on the chaoticity and on the main
UPO-associated signatures is displayed in Fig.~\ref{fig:robustness}. In
panel~(a), the median finite-time Lyapunov exponent is shown as a function of
$N$ for the three models. The original homogeneous model with
$\epsilon=0$ displays the largest median FTLE for every environment size
considered. Introducing the central bias substantially reduces the FTLE.
When environmental inhomogeneity is subsequently introduced, the FTLE
increases again relative to the centrally biased homogeneous model for
$N=6,8,$ and $10$, although it remains below the values of the original
homogeneous model.

Panel~(b) shows the normalized Floquet instability of the selected UPOs.
Unlike the FTLE, no simple dependence on the bias configuration is observed.
The centrally biased model gives the largest mean Floquet instability for
$N=4,6,8$, whereas the original homogeneous model gives the largest
values for $N=2$ and $N=10$. The effect of the bias on the local
instability of the UPOs therefore depends on the environment size and does
not follow a monotonic trend.

Panel~(c) compares the number of UPO families showing statistically
significant localization and recurrence signatures. For the full-window tube
localization, significant behavior is found for 18/20 families in the
homogeneous model and for 20/20 families in both modified models. At late
times, the corresponding numbers are 14/20, 17/20, and 20/20. For the
return fidelity enhancement over the first five periods, significant results
are obtained for 19/20, 17/20, and 19/20 families, respectively. Finally,
the recurrence-timing criterion is satisfied by all 20 UPO families in the
three models.

The central bias therefore does not produce a systematic increase or
decrease in the number of UPO families displaying these signatures. The
environmentally inhomogeneous model, however, gives significant results for
almost all UPO families for each of the four diagnostics considered here. The UPO-associated signatures
therefore remain highly robust after the environmental homogeneity and the
corresponding Gram conservation are lost.

Taken together, these results show that the UPO-associated dynamical
signatures identified in the homogeneous reference model persist after a
finite central bias is introduced and also when different biases are assigned
to the environmental TLSs. In the latter case, the Gram quantities are no
longer conserved, yet the spatial and temporal signatures remain
highly robust across the complete UPO catalogue. The observed
UPO-associated memory is therefore not restricted to the exact homogeneous
structure of the original model.

\section{Conclusions}\label{sec:conclusions}

In this work, it is studied whether unstable periodic orbits can organize
the dynamics of a nonlinear system--environment model composed of a central
two-level system coupled to a finite environment of additional TLSs. Unlike
conventional quantum-scar studies based on the eigenstates of a fixed linear
Hamiltonian, the present system evolves according to nonlinear,
state-dependent equations. The analysis therefore focuses on dynamical
signatures associated with UPOs rather than on eigenstate localization.

The study is organized around three complementary analyses. First, the chaotic dynamics and UPO-associated localization of the homogeneous reference model are characterized while varying the number of environmental TLSs, $N=2,4,6,8,$ and $10$, at fixed energy density. Second, the dependence on energy is examined by keeping $N=10$ fixed and repeating the analysis over different energy shells. Finally, the robustness of the observed behavior against changes in the symmetry of the model is tested by first introducing an energetic asymmetry in the central TLS and then an inhomogeneous environment in which the environmental TLSs have different energetic biases.

For the homogeneous reference model, finite-time Lyapunov exponents confirm
that the selected energy regions are chaotic, while the periodic-orbit
search identifies unstable recurrent trajectories embedded in this
background. Their local stability is characterized through fixed-sector
Floquet analysis, and their influence on nearby chaotic trajectories is
examined through spatial and temporal observables. In particular, the occupation of the complete UPO tube, its persistence at late times, the enhancement of return fidelity, and the timing of recurrent returns are evaluated. Taken together, these quantities show that trajectories initialized
close to a UPO retain a measurable association with both its spatial path
and its characteristic recurrence time after the immediate effect of the
initial condition has been excluded. This behavior is referred to as 
UPO-associated dynamical localization.

In addition, the energy dependence of this behavior is studied for
$N=10$. Increasing the energy produces a pronounced increase in the
chaoticity of the surrounding dynamics, while the strength of the
UPO-associated spatial and return fidelity signatures generally decreases.
The local Floquet instability does not follow a simple monotonic dependence
on energy. By contrast, the timing of the recurrent returns remains closely
linked to the UPO period throughout the energy range considered. These
results indicate that the strength of UPO-associated localization depends on
the dynamical regime, while the characteristic recurrence times of the UPOs
remain detectable even as the surrounding dynamics becomes more chaotic.

Finally, the robustness analysis shows that the observed UPO-associated
signatures are not restricted to the exact symmetry of the reference model.
Introducing a finite bias in the central TLS reduces the chaoticity of the
system while preserving the localization and recurrence signatures. When
the environmental TLSs are subsequently made nonidentical, the chaoticity
increases again for some environment sizes, although it remains below that
of the original homogeneous model. The UPO-associated signatures also
persist in this case, even though the Gram invariants of the homogeneous
environment are no longer conserved. The UPO-associated organization of
nearby trajectories therefore survives both changes in the central system
and the loss of environmental homogeneity, even when these modifications
substantially alter the surrounding chaotic dynamics.

The three analyses lead to a common picture. UPOs remain dynamically
relevant across changes in environment size, energy, and model symmetry,
although the magnitude of the individual localization measures depends on
the surrounding chaotic regime. The results do not suggest a simple
one-to-one relation between chaoticity, Floquet instability, and localization
strength. Instead, spatial localization, recurrence amplitude, and recurrence
timing provide complementary information about how unstable periodic motion
influences nearby chaotic trajectories.

More generally, the model considered here provides a convenient nonlinear
chaotic framework in which UPOs can be studied while independently varying
system size, energy, symmetry, and system--environment parameters. Its
system--environment structure also makes it possible to examine how recurrent
dynamics changes when a central subsystem interacts with an increasingly
complex or inhomogeneous environment. In this setting, a scar-like
dynamical analysis can be performed by testing whether trajectories
initialized close to UPOs retain measurable spatial and temporal signatures
of the underlying periodic motion. Beyond the specific TLS realization, the
dynamical-systems framework employed here is relevant to a broad class of
nonlinear problems in applied mathematics and physics in which unstable
recurrent structures organize chaotic motion. The present analysis
therefore connects nonlinear system--environment dynamics with the broader
use of unstable periodic orbits as organizing structures of complex chaotic
dynamics.

\appendix
\newcommand{\appsection}[2]{%
    \refstepcounter{section}%
    \section*{Appendix~\thesection:\ #1}%
    \label{#2}%
}

\appsection
{Collective structure and Gram invariants}
{app:collective_structure}

For a homogeneous environment, all environmental TLSs have the same
dynamical parameters,
\begin{equation}
    \delta_i=\delta_{\mathrm e},
    \qquad
    \epsilon_i=\epsilon_{\mathrm e},
    \qquad
    \Lambda_i=\Lambda .
\end{equation}
This does not imply that the environmental TLSs have identical instantaneous
states. Instead, it means that all of them follow the same equations of
motion and interact in the same way with the central TLS.

The collective environmental Bloch vector is defined as
\begin{equation}
    \bm{S}
    =
    \sum_{i=1}^{N}\bm{r}_i
    =
    (S_x,S_y,S_z).
    \label{eq:app_collective_spin}
\end{equation}
Using this quantity, the Hamiltonian of the homogeneous model becomes
\begin{equation}
    H
    =
    -2\delta x_0
    +2\epsilon z_0
    -2\delta_{\mathrm e}S_x
    +2\epsilon_{\mathrm e}S_z
    +\Lambda z_0S_z.
    \label{eq:app_collective_hamiltonian}
\end{equation}
The interaction between the central TLS and the environment therefore depends
only on the collective environmental coordinate $S_z$. Consequently, the
evolution of the central TLS is determined by collective environmental
variables rather than by the individual environmental Bloch vectors
separately.

The same homogeneity also imposes an additional invariant structure on the
environmental dynamics. Since every environmental Bloch vector evolves under
the same instantaneous effective field generated by the central TLS, all
environmental vectors undergo the same time-dependent rotation. Their
relative geometry is therefore preserved during the evolution.

For any two environmental TLSs, the pairwise Gram quantity is defined as
\begin{equation}
    G_{ij}
    =
    \bm{r}_i\cdot\bm{r}_j.
\end{equation}
Because all environmental TLSs obey the same equations of motion with the
same time-dependent central coordinate $z_0(t)$, differentiation of
$G_{ij}$ gives
\begin{equation}
    \frac{dG_{ij}}{dt}
    =
    \dot{\bm{r}}_i\cdot\bm{r}_j
    +
    \bm{r}_i\cdot\dot{\bm{r}}_j
    =
    0.
\end{equation}
Therefore,
\begin{equation}
    G_{ij}(t)=G_{ij}(0),
    \label{eq:app_gram_conservation}
\end{equation}
and the relative geometry of the environmental Bloch vectors remains
unchanged during the evolution. The conserved values of $G_{ij}$ therefore
define invariant sectors of the dynamics.

This invariant structure is used in the main analysis when constructing
fixed-sector perturbations for the Floquet calculation and when comparing
near-UPO trajectories with reference trajectories belonging to the same
dynamically accessible sector.

\appsection
{Accessible energy ranges}
{app:energy_ranges}

For the homogeneous model considered in the main analysis, the parameters are
\begin{equation}
    \delta=\delta_i=1,
    \qquad
    \epsilon=\epsilon_i=0,
    \qquad
    \Lambda_i=1.
\end{equation}
The Hamiltonian can therefore be written as
\begin{equation}
    H
    =
    -2\sqrt{1-z_0^2}\cos\Phi
    + \sum_{i=1}^{N}
    \left[
        -2\sqrt{1-z_i^2}\cos\phi_i
    \right]+ z_0
    \sum_{i=1}^{N}z_i.
\end{equation}

The accessible energy bounds can be obtained by maximizing and minimizing
this Hamiltonian over all allowed Bloch-sphere coordinates. 
Since all population coordinates are restricted to the interval
$-1\leq z_k\leq1$, maximizing the Hamiltonian over the allowed Bloch-sphere
coordinates gives
\begin{equation}
    E_{\max}(N)
    =
    \begin{cases}
        2+2N, & N\leq4,\\[4pt]
        \sqrt{5(N^2+4)}, & N\geq4.
    \end{cases}
    \label{eq:app_energy_bound}
\end{equation}

On the other hand, the symmetry of the Hamiltonian gives the corresponding lower bound,
\begin{equation}
    E_{\min}(N)
    =
    -E_{\max}(N).
\end{equation}

Therefore, for the five environment sizes considered in this work, the minimum and maximum energies are listed in
Table~\ref{tab:app_energy_bounds}.

\begin{table}[H]
    \centering
    \small
    \begin{tabular}{ccc}
        \hline
        $N$ &
        $E_{\min}(N)$ &
        $E_{\max}(N)$ \\
        \hline
        2  & -6.0000  & 6.0000  \\
        4  & -10.0000 & 10.0000 \\
        6  & -14.1421 & 14.1421 \\
        8  & -18.4391 & 18.4391 \\
        10 & -22.8035 & 22.8035 \\
        \hline
    \end{tabular}
    \caption{\justifying
    Minimum and maximum energies accessible to the homogeneous model for the
    five numbers of environmental TLSs considered in this work.
    }
    \label{tab:app_energy_bounds}
\end{table}

To compare energies across systems with different numbers of environmental
TLSs, the normalized energy coordinate is defined as
\begin{equation}
    \xi_N(\mathcal{E})
    =
    \frac{
        \mathcal{E}-E_{\min}(N)
    }{
        E_{\max}(N)-E_{\min}(N)
    }.
    \label{eq:app_normalized_energy}
\end{equation}
With this definition, $\xi_N=0$ corresponds to the minimum accessible energy
and $\xi_N=1$ to the maximum.

The reference energy originally selected for the $N=2$ system is
\begin{equation}
    \mathcal{E}_2=2.51135217.
\end{equation}
Its normalized energy coordinate is
\begin{equation}
    \xi_2(\mathcal{E}_2)=0.7093.
\end{equation}
Since the $N=2$ system contains three TLSs in total, this reference energy
corresponds to the energy per TLS
\begin{equation}
    e
    =
    \frac{\mathcal{E}_2}{3}
    =
    0.83711739.
    \label{eq:app_reference_energy_density}
\end{equation}

This value of $e$ is kept fixed in the main $N$-dependent analysis. The total
energy used for each environment size is therefore
\begin{equation}
    \mathcal{E}_N
    =
    (N+1)e.
    \label{eq:app_target_energy}
\end{equation}
The corresponding total energies and their normalized positions are

\begin{table}[H]
    \centering
    \small
    \begin{tabular}{ccc}
        \hline
        $N$ &
        $\mathcal{E}_N$ &
        $\xi_N(\mathcal{E}_N)$ \\
        \hline
        2  & 2.5114 & 0.7093 \\
        4  & 4.1856 & 0.7093 \\
        6  & 5.8598 & 0.7072 \\
        8  & 7.5341 & 0.7043 \\
        10 & 9.2083 & 0.7019 \\
        \hline
    \end{tabular}
    \caption{\justifying
    Total energies used in the main $N$-dependent analysis and their
    corresponding normalized energy coordinates. All values are obtained by
    keeping the energy per TLS fixed at $e=0.83711739$.
    }
    \label{tab:app_selected_energies}
\end{table}

The values of $\xi_N(\mathcal{E}_N)$ vary only from $0.7093$ to $0.7019$.
Thus, keeping the energy per TLS fixed at
\begin{equation}
    e=0.83711739
\end{equation}
selects comparable regions of the accessible energy interval for all five
environment sizes considered in the main $N$-dependent analysis.

The energy-dependent analysis of Sec.~\ref{sec:energy_dependence} instead
keeps the environment size fixed at $N=10$. For this system, the reference
energy
\begin{equation}
    \mathcal{E}_{10}=9.2083
\end{equation}
corresponds to the normalized energy coordinate
\begin{equation}
    \xi_{10}(\mathcal{E}_{10})=0.7019.
\end{equation}
This value is taken as the central point of the energy scan. Four additional
energy shells are selected using two values below and two above the reference,
with a constant spacing
\begin{equation}
    \Delta\xi_{10}=0.075.
\end{equation}
The five normalized energy coordinates are therefore
\begin{equation}
    \xi_{10}
    =
    0.5519,\;
    0.6269,\;
    0.7019,\;
    0.7769,\;
    0.8519.
\end{equation}
Using Eq.~\eqref{eq:app_normalized_energy}, these values correspond to the
energy densities
\begin{equation}
    e
    =
    0.2152,\;
    0.5262,\;
    0.8371,\;
    1.1481,\;
    1.4590.
    \label{eq:app_energy_scan_density}
\end{equation}
These are the five energy densities used in
Sec.~\ref{sec:energy_dependence}.

\appsection
{Search and numerical validation of unstable periodic orbits}
{app:upo_search}

Trajectories that return \emph{close} to a previous state are therefore searched for first, and these approximate recurrences are subsequently refined numerically. The search is based on a Poincar\'e section
\cite{Cvitanovic1991}.

A Poincar\'e section reduces a continuous trajectory to the sequence of
points at which it crosses a chosen surface in phase space. This makes close
returns considerably easier to identify. The section
\begin{equation}
    y_0=0,
    \qquad
    x_0>0,
    \label{eq:poincare_section}
\end{equation}
is used, with crossings taken in a fixed direction. Since
\begin{equation}
    y_0=\sqrt{1-z_0^2}\sin\Phi,
\end{equation}
the conditions $y_0=0$ and $x_0>0$ select the central phase
$\Phi=0$ and remove the second crossing corresponding to $\Phi=\pi$.

Long trajectories are generated from different initial conditions at the
target energy. Each time a trajectory crosses the Poincar\'e section, the
complete state of the central and environmental TLSs is recorded. The
condition in Eq.~\eqref{eq:poincare_section} is therefore imposed only on
the central TLS; the environmental variables are not fixed and take the
values generated dynamically at each crossing.

Pairs of section crossings that are very close to one another but separated
by a finite time provide approximate periodic-orbit candidates. For each
candidate, both the initial state and the return time $T$ are then adjusted
numerically until the periodicity condition
\begin{equation}
    \bm{R}(T)-\bm{R}(0)
    \simeq
    \bm{0}
\end{equation}
is satisfied while keeping the total energy fixed. In other words, the close
recurrence provides an initial estimate of the periodic orbit, and the
numerical refinement progressively modifies that estimate until the complete
state closes after one period.

Once a periodic solution has been found, it is integrated again using
stricter numerical tolerances. It is then independently verified that the final state reproduces the initial state and that the energy remains constant throughout one complete period. This additional integration ensures that the
periodic solution is not an artifact of the numerical optimization. The
Python implementation used for the recurrence search, numerical refinement,
and validation is provided.

\section*{Acknowledgements}
D. M.-G. gratefully acknowledges Pedro Bargueño and Salvador Miret-Artés for their valuable discussions, guidance, and support during the development of the initial model on which this work is based.

\section*{Data availability}
All Python code used in this work, together with instructions to reproduce the reported results, are available at \url{https://github.com/daniel-martinez-gil/UPOs}.
\bibliographystyle{unsrt}
\bibliography{referencias.bib}
\end{document}